\documentclass[universe,article,accept,pdftex,moreauthors]{mdpi_mod} 

\newcommand{\Msun}{\ensuremath{\mathrm{\rm M}_\odot}}
\newcommand{\Mbh}{\ensuremath{M_\mathrm{\rm BH}}}
\newcommand{\Mnsc}{$M_\mathrm{\rm NSC}$}
\newcommand{\Mstar}{$M_{\star}$}

\newcommand{\Lsun}{\ensuremath{\mathrm{\rm L}_\odot}}

\newcommand{\Rsoi}{$r_{\rm SOI}$}

\newcommand{\ml}{\ensuremath{M/L}}

\newcommand{\hst}{\emph{HST}}

\newcommand{\kms}{km~s$^{-1}$}

\newcommand{\trelax}{t$\rm _{relax}$}
\mathchardef\mhyphen="2D
\newcommand{\hsim}{{\tt HSIM}}

\newcommand{\umu}{\mu}

\usepackage{graphicx}	
\usepackage{amsmath}	
\usepackage{multirow}
\usepackage{amssymb}
\DeclareUnicodeCharacter{2605}{\star}
\DeclareUnicodeCharacter{03B1}{$\alpha$}   
\DeclareUnicodeCharacter{03B2}{$\beta$}    
\DeclareUnicodeCharacter{03B3}{$\gamma$}   
\DeclareUnicodeCharacter{03B4}{$\delta$}   
\DeclareUnicodeCharacter{03B5}{$\epsilon$} 
\DeclareUnicodeCharacter{03B6}{$\zeta$}    
\DeclareUnicodeCharacter{03B7}{$\eta$}     
\DeclareUnicodeCharacter{03B8}{$\theta$}   
\DeclareUnicodeCharacter{03BB}{$\lambda$}  
\DeclareUnicodeCharacter{03C0}{$\pi$}      
\DeclareUnicodeCharacter{03C3}{$\sigma$}   
\DeclareUnicodeCharacter{03C6}{$\phi$}     
\DeclareUnicodeCharacter{03C9}{$\omega$}   
\DeclareUnicodeCharacter{22C6}{$\star$}   

\usepackage[utf8]{inputenc}
\firstpage{1} 
\pubvolume{1}
\issuenum{1}
\articlenumber{0}
\pubyear{2025}
\copyrightyear{2025}
\datereceived{ } 
\daterevised{ } 
\dateaccepted{ } 
\datepublished{ } 
\hreflink{https://doi.org/} 

\Title{Detecting Intermediate-Mass Black Holes out to 20 Mpc with ELT/HARMONI: The Case of FCC 119}

\TitleCitation{Simulating IMBH measurement in FCC 119 with HARMONI IFS}

\Author{Hai N. Ngo $^{1,4}$*\orcidA{}, Dieu D. Nguyen $^{2}$\orcidB{}, Tinh Q. T. Le $^{3}$\orcidC{}, Tien T. H. Ho$^{1,4}$\orcidD{}, Truong N. Nguyen $^{1,4}$\orcidE{} and Trung H. Dang $^{1,4}$\orcidF{}}

\AuthorNames{Hai N. Ngo, Dieu D. Nguyen, Tinh T. Q Le, Tien T. H. Ho, Truong N. Nguyen and Trung H. Dang}

\isAPAStyle{%
       \AuthorCitation{Ngo, H.}
         }{%
        \isChicagoStyle{%
        \AuthorCitation{Ngo, H.}
        }{
        \AuthorCitation{Ngo, H.}
        }
}

\address{%
$^{1}$ \quad Faculty of Physics -- Engineering Physics, University of Science, Vietnam National University, Ho Chi Minh City, Vietnam; \\
$^{2}$ \quad Department of Astronomy, University of Michigan, 1085 South University Avenue, Ann Arbor, MI 48109, USA; nddieuphys@gmail.com \\
$^{3}$ \quad International Centre for Interdisciplinary Science and Education, 07 Science Avenue, Ghenh Rang, 55121 Quy Nhon, Vietnam; \\
$^{4}$ \quad Vietnam National University, Ho Chi Minh City, Vietnam; 
}

\corres{Correspondence: hai10hoalk@gmail.com;}

\abstract{Intermediate-mass black holes (IMBHs; $\Mbh \approx 10^{3-5}\Msun$) play a critical role in understanding the formation of supermassive black holes in the early universe. In this study, we expand on Nguyen et al. simulated measurements of IMBH masses using stellar kinematics, which will be observed with the High Angular Resolution Monolithic Optical and Near-infrared Integral (HARMONI) field spectrograph on the Extremely Large Telescope (ELT) up to the distance of 20 Mpc. Our sample focuses on both the Virgo Cluster in the northern sky and the Fornax Cluster in the southern sky. We begin by identifying dwarf galaxies hosting nuclear star clusters, which are thought to be nurseries for IMBHs in the local universe. As a case study, we conduct simulations for FCC 119, the second faintest dwarf galaxies in the Fornax Cluster at 20 Mpc, which is also fainter than most of Virgo Cluster members. We use the galaxy’s surface brightness profile from Hubble Space Telescope (HST) imaging, combined with an assumed synthetic spectrum, to create mock observations with the {\tt HSIM} simulator and Jeans Anisotropic Models (JAM). These mock HARMONI datacubes are analyzed as if they were real observations, employing JAM within a Bayesian framework to infer IMBH masses and their associated uncertainties. We find that ELT/HARMONI can detect the stellar kinematic signature of an IMBH and accurately measure its mass for $\Mbh \gtrsim 10^5~\Msun$ out to distances of $\sim$20 Mpc.}

\begin{document}
\section{Introduction}

Intermediate-mass black holes (IMBHs; $M_{\rm BH}\approx10^{3-5}\Msun$) in dwarf galaxies ($M_{\star}\lesssim10^{9}\Msun$) represent a crucial but largely missing component of the black hole (BH) mass spectrum \citep{Greene12, Mezcua17}. Studying the correlations between IMBH mass and the macroscopic properties of their host galaxies—such as central stellar velocity dispersion ($\sigma_\star$) \citep{Ferrarese00, Gebhardt00} or bulge stellar mass ($M_{\rm bulge}$) \citep{kormendy2001supermassive}—provides valuable insights when compared to their supermassive black hole (SMBH; $M_{\rm BH}\approx10^{6-10}\Msun$) counterparts in more massive galaxies ($M_{\star}\gtrsim10^{10}\Msun$). Additionally, determining the occupation fraction—the proportion of dwarf galaxies hosting IMBHs—can constrain potential formation pathways of these black holes in the early universe \citep{Greene20}. Despite extensive searches \citep{Greene20}, the presence of IMBHs remains elusive, and their origins continue to be an open question in astrophysics \citep{Inayoshi20, Neumayer20}.

Despite the fact that dynamical methods using stellar or gas motions in the centers of nearby dwarf galaxies are among the most precise techniques for searching for strong evidence of IMBHs \citep{Ahn18, denBrok15, Davis20, Nguyen17conf, Nguyen19conf, Nguyen17, Nguyen18, Nguyen19, Nguyen20, Nguyen21, Nguyen22, Nguyen2025a, Thater20, Thater2022, Thater2023, Voggel2018}, evidence for IMBHs remains limited to a handful of candidates. Several strong dynamical cases are found in some nearby massive globular clusters, such as Omega Centauri \citep[\Mbh $\gtrsim 8,200\ \Msun$;][]{Haberle24}, 47~Tucanae \citep[\Mbh $\gtrsim 578\ \Msun$;][]{kiziltan17, croce24}. Furthermore, Dark Energy Spectroscopic Instrument (DESI) survey catalogued $\sim$300 IMBH candidates with $\Mbh \lesssim 10^6~\Msun$\citep {pucha25}. However, definitive proof of their dynamical mass is still limited.

Dynamical confirmation remains observationally challenging due to the insufficient spatial resolution of current facilities \citep{Seth10, Nguyen17} to resolve the black hole’s sphere of influence (SOI), defined as $r_{\rm SOI} = GM_{\rm BH}/\sigma_\star^2$, where $G$ is the gravitational constant. For example, for an IMBH with $\Mbh \approx 10^5$ M$_\odot$ and $\sigma_\star \approx 30$ \kms\ located at a distance just beyond the Local Group ($D \approx 3.5$~Mpc) the corresponding \Rsoi $\approx 0\farcs028$ \citep[e.g., see Eq.~1 of][]{Nguyen23}, which is smaller than the diffraction limits of current 8–10 meter class ground-based telescopes equipped with adaptive optics (AO), which achieve point-spread functions (PSFs) with typical full widths at half maximum (FWHM) in the range of $\rm FWHM_{PSF} \approx 0\farcs05$–$0\farcs1$. This limitation raises the question of whether the current lack of strong observational evidence for IMBHs reflects their true absence, or simply the restrictions imposed by existing instrumentation.

The next generation of giant ground-based telescopes may provide a solution to this problem. In particular, the Extremely Large Telescope (ELT) is poised to overcome current resolution challenges with its 39 meter primary mirror, equipped with the Multi-AO Imaging Camera for Deep Observations (MICADO; $\rm FWHM_{PSF} \approx 0\farcs01$) imager \citep{Davies10, Davies21} and the High Angular Resolution Monolithic Optical and Near-infrared Integral (HARMONI; $\rm FWHM_{PSF} \approx 0\farcs012$) field spectrograph, which also offers a high-spectral resolution ($R \approx 17{,}400$) \citep{Thatte16, Thatte20}. However, to locate potential IMBHs, we must first identify suitable targets. According to \citep{Nguyen2025b}, bright nuclear star clusters (NSCs) within 10 Mpc are promising hosts of IMBHs. These IMBHs can be dynamically detected if their masses are at least $\sim$0.5\% of their NSC mass ($M_{\rm NSC}$), although the effects of dynamical mass segregation must be carefully considered before drawing strong conclusions.

In this study, we extend the distance limit of the HARMONI IMBH survey \citep{Nguyen2025b} to include the Virgo ($D \approx 16.5$ Mpc) and Fornax ($D \approx 20$ Mpc) clusters by simulating $\Mbh$ measurements for FCC~119, the second faintest nucleated member of the Fornax Cluster and significantly fainter than nearly all members of the Virgo Cluster. These measurements are derived from high-spatial-resolution integral-field unit (IFU) stellar kinematics extracted from mock HARMONI data, generated using the HARMONI Simulator ({\tt HSIM}\footnote{v3.11: \url{https://github.com/HARMONI-ELT/HSIM}}) \citep{Zieleniewski15}. We also investigate the observational limits in terms of the minimum NSC central surface brightness (SB), IMBH mass, and on source exposure time that HARMONI can effectively detect.

We describe the criteria used to finalize the extended HARMONI IMBH sample and its properties, as well as the characteristics of the Virgo and Fornax Clusters, which contain a significant portion of our sample, in Section \ref{sec:sample-selection}. In Section \ref{Sec:Stellar-mass-model}, we present the stellar mass model of FCC~119. Section \ref{sec:HARMONI-simulation} provides a detailed description of our HARMONI IFS simulations. Finally, we discuss the results in Section \ref{sec:discussion} and summarize our findings in Section \ref{sec:Conclusions}.

We adopt a flat Universe with a Hubble constant of $H_0 \approx 70$ km s$^{-1}$ Mpc$^{-1}$, a matter density of $\Omega_{\rm m,0} \approx 0.3$, and a dark energy density of $\Omega_{\Lambda,0} \approx 0.7$, which give a physical scale\footnote{Ned Wright Cosmology Calculator: \url{https://www.astro.ucla.edu/~wright/CosmoCalc.html}} of 100 pc arcsecond$^{-1}$. All photometric measurements in this study are reported in the AB magnitude system \citep{Oke1974} and corrected for foreground extinction \citep{Schlafly11} using a interstellar extinction law \citep{Cardelli89}. Furthermore, all kinematic maps are aligned such that the galaxy’s major and minor axes correspond to the horizontal and vertical directions, respectively.

\section{SAMPLE SELECTION}\label{sec:sample-selection}

\subsection{Sample selection}\label{sec:select-process}

We began with the 2MASS Ks-band photometry \citep{Jarrett2003, Skrutskie06} from the 2MASS Redshift Survey (2MRS) \citep{Huchra12} as our parent catalog. Stellar masses were estimated from $K_s$-band luminosities by adopting a constant $\ml_K \approx 1$ (\Msun/\Lsun) \citep{Nguyen18} and using distances from the Updated Nearby Galaxy Catalog \citep{Karachentsev13}. The virial stellar velocity dispersion was then calculated following the relation $\sigma_{\rm vir}^2 = GM/(5R_e)$ \citep{cappellari06}. We applied the following selection criteria to define our candidate sample:

\begin{itemize}
    \item Virial stellar velocity dispersion $\sigma_{\rm vir} \leq 75$ km s$^{-1}$. This threshold is set based on the extrapolated $M_{\rm BH}$–$\sigma_\star$ relation \citep{Greene20}, corresponding to the IMBH regime. Moreover, systems with lower dispersions yield larger projected angular sizes for the IMBH $R_{\rm SOI}$, enhancing the prospects for resolving their kinematics.
    
    \item Distance range of $10 < D \lesssim 20$ Mpc. This extends the selection beyond the nearer targets included in the HARMONI IMBH sample \citep{Nguyen2025b}.

    \item Targets lie within the ELT’s observability, defined as $|\delta + 24^\circ| < 45^\circ$, and outside the Galactic plane (zone of avoidance) with Galactic latitude of $|b| > 8^\circ$.
\end{itemize}

The resulting candidate list was then cross-checked against existing compilations of nucleated dwarf galaxies from both photometric and spectroscopic NSC surveys \citep{Boker02, Rossa06, Ho09, Cappellari11, Neumayer12, Graham12, Georgiev14, Georgiev16, Spengler17, Graham19b, Graham19a, Pechetti20, Fahrion2021, Baldassare22, Fahrion22, Hoyer23}. Finally, we visually inspected the galaxies using imaging data from the available data archives to confirm the presence of NSCs (Figure \ref{fig:sample}).

\begin{table}
\caption{Selected criteria for dwarf galaxies.}
\centering
\begin{tabular}{rl}
\hline
ELT Observability: & $|\delta+24^\circ|<45^\circ$ \\
Galaxy zone of avoidance: & $|b|>8^\circ$ \\
Central stellar-velocity dispersion: & $\sigma_{\star,\rm c} < 75$ \kms \\
\hline
\label{tab:parent_criteria}
\end{tabular}
\parbox[t]{1.0\textwidth}{\textit{Notes:}$\delta$: the Declination, $b$: the Galactic latitude.}
\end{table}

\begin{table}
\caption{Properties of the selected sample}
\centering
\begin{tabular}{rl}
\hline
Distant range: & $10< D\lesssim20$ Mpc \\
Galaxies $K_s$-band absolute magnitude: & $-23.7 \lesssim M_K/{\rm mag} \lesssim -17.4$ \\
Galaxy stellar mass range: & $10^8$ $\lesssim$ M$_{\star,\rm gal}/\Msun\lesssim 5 \times 10^{10}$ \\
Effective : & $R_{\rm e, gal} \lesssim 7$ kpc \\
NSC mass range: & $1.6\times10^5 < M_{\rm NSC}/\Msun < 2.3\times10^8$ \\
NSC effective radius: & $0.5 < R_{\rm e} < 62$ pc\\
\hline
\label{tab:selected_sample}
\end{tabular}
\end{table}

\begin{table*}
\footnotesize
\begin{adjustwidth}{-\extralength}{0cm}
\caption{Full list and properties of our extended HARMONI IMBH sample hosting NSCs within the distant range of 10--20 Mpc}
\centering
\begin{tabular}{cccccccccccc}
\hline\hline
No.&Object&RA&Decl.&$D$&Hubble&$M_{\star, \rm gal}$&$R_{\rm e,gal}$&$M_{\rm NSC}$&$R_{\rm e,NSC}$& $\sigma_{\star,\rm NSC}$& Ref. \\
 & name & (h:m:s) & (d:m:s) &(Mpc)&type&(\Msun)&(kpc)&(\Msun)&(pc)&(\kms)& ~ \\ 
(1) & (2) & (3) & (4) & (5) & (6) & (7) & (8) & (9) & (10) & (11) & (12)\\\hline
1  & ESO301-IG11 & 03:23:54.21 & $-$37:30:33.0 & 19.94 & 9.9    & 9.61  & 2.11 & 5.79 & 3.7  & 28.9 & \citep{Georgiev16} \\
2  & IC1933      & 03:25:39.97 & $-$52:47:07.5 & 18.98 & 6.1    & 8.35  & 1.85 & 5.21 & 6.2  & 42.7 & \citep{Georgiev16} \\
3  & MCG-1-03-85 & 01:05:04.88 & $-$06:12:44.6 & 11.38 & 7.0    & 9.66  & 4.04 & 6.40 & 2.9  & 42.0 & \citep{Georgiev16,Baldassare22} \\
4  & NGC0428     & 01:12:55.75 & +00:58:53.5   & 16.54 & 8.6    & 9.54  & 2.63 & 6.41 & 1.2  & 56.9 & \citep{Georgiev16} \\
5  & NGC0864     & 02:15:27.64 & +06:00:09.4   & 19.41 & 5.1    & 10.10 & 0.21 & 7.86 & 2.7  & 26.9 & \citep{Georgiev16,Ho09} \\
6  & NGC1042     & 02:40:23.97 & $-$08:26:00.7 & 19.59 & 6.0    & 9.32  & 4.59 & 6.33 & 1.3  & 65.3 & \citep{Georgiev16} \\
7  & NGC1073     & 02:43:40.52 & +01:22:34.0   & 13.80 & 5.3    & 9.69  & 0.16 & 6.53 & 0.5  & 24.8 & \citep{Georgiev16,Ho09} \\
8  & NGC1493     & 03:57:27.46 & $-$46:12:38.6 & 15.54 & 6.0    & 9.77  & 4.85 & 6.55 & 3.6  & 51.2 & \citep{Georgiev16} \\
9  & NGC1518     & 04:06:49.82 & $-$21:10:23.5 & 13.50 & 8.2    & 9.10  & 0.83 & 5.35 & 3.4  & 36.6 & \citep{Georgiev16} \\
10 & NGC1559     & 04:17:35.77 & $-$62:47:01.2 & 15.21 & 5.9    & 9.77  & 3.50 & 6.66 & 1.5  & 72.6 & \citep{Georgiev16,Baldassare22,Sun20} \\
11 & NGC2835     & 09:17:52.91 & $-$22:21:16.8 & 10.86 & 5.0    & 9.70  & 2.80 & 6.61 & 3.4  & 70.7 & \citep{Georgiev16,Baldassare22,Sun20} \\
12 & NGC3346     & 10:43:38.91 & +14:52:18.9   & 17.84 & 5.9    & 9.51  & 3.34 & 6.37 & 1.1  & 49.6 & \citep{Georgiev16} \\
13 & NGC3423     & 10:51:14.33 & +05:50:24.1   & 11.27 & 6.0    & 9.45  & 0.12 & 6.29 & 2.2  & 54.6 & \citep{Georgiev16} \\
14 & NGC3455     & 10:54:31.08 & +17:17:04.6   & 15.79 & 3.7    & 10.57 & 1.44 & 6.81 & 2.2  & 46.2 & \citep{Georgiev16} \\
15 & NGC3666     & 11:24:26.07 & +11:20:32.0   & 19.32 & 5.2    & 9.94  & 0.10 & 7.81 & 9.6  & 60.6 & \citep{Georgiev16,Ho09} \\
16 & NGC4212     & 12:15:39.36 & +13:54:05.4   & 17.30 & 4.9    & 10.40 & 4.56 & 6.70 & 12.3 & 61.0 & \citep{Georgiev16,Graham19b} \\
17 & NGC4487     & 12:31:04.46 & $-$08:03:14.1 & 19.59 & 6.0    & 10.38 & 3.50 & 6.93 & 1.1  & 51.0 & \citep{Georgiev16} \\
18 & NGC4496A    & 12:31:39.21 & +03:56:22.1   & 15.49 & 7.6    & 9.65  & 3.10 & 6.23 & 0.7  & 74.9 & \citep{Georgiev16,Ho09,Sun20} \\
19 & NGC4504     & 12:32:17.41 & $-$07:33:48.9 & 14.51 & 6.0    & 10.54 & 2.10 & 6.85 & 3.9  & 53.0 & \citep{Georgiev16} \\
20 & NGC4517     & 12:32:45.59 & +00:06:54.1   & 10.67 & 6.0    & 10.08 & 6.50 & 5.85 & 2.3  & 43.8 & \citep{Georgiev16,Ho09} \\
21 & NGC4592     & 12:39:18.74 & $-$00:31:55.0 & 15.21 & 8.0    & 10.28 & 1.84 & 5.80 & 1.1  & 42.6 & \citep{Georgiev16} \\
22 & NGC4635     & 12:42:39.23 & +19:56:43.7   & 13.54 & 6.6    & 9.01  & 1.40 & 6.79 & 4.3  & 34.7 & \citep{Georgiev16} \\
23 & NGC4771     & 12:53:21.26 & +01:16:09.5   & 16.07 & 6.2    & 10.32 & 2.73 & 6.63 & 1.9  & 64.4 & \citep{Georgiev16} \\
24 & NGC4900     & 13:00:39.26 & +02:30:02.7   & 13.73 & 5.2    & 10.12 & 2.59 & 7.24 & 3.1  & 67.1 & \citep{Georgiev16} \\
25 & NGC4904     & 13:00:58.64 & $-$00:01:39.4 & 16.94 & 5.8    & 9.92  & 2.06 & 6.23 & 2.5  & 62.5 & \citep{Georgiev16} \\
26 & NGC5054     & 13:16:58.49 & $-$16:38:05.5 & 18.62 & 4.2    & 10.43 & 0.11 & 7.02 & 8.3  & 57.4 & \citep{Georgiev16,Baldassare22} \\
27 & NGC5300     & 13:48:16.04 & +03:57:03.1   & 16.74 & 5.2    & 9.97  & 2.68 & 7.21 & 20.6 & 54.1 & \citep{Georgiev16} \\
28 & NGC5334     & 13:52:54.48 & $-$01:06:52.0 & 19.63 & 5.2    & 9.79  & 2.96 & 6.85 & 11.9 & 49.5 & \citep{Georgiev16} \\
29 & NGC7090     & 21:36:28.81 & $-$54:33:25.3 & 11.86 & 5.0    & 9.64  & 4.56 & 6.48 & 1.1  & 54.3 & \citep{Georgiev16} \\
30 & NGC7424     & 22:57:18.37 & $-$41:04:14.1 & 11.48 & 6.0    & 9.75  & --   & 6.00 & 6.8  & 15.6 & \citep{Georgiev16,Walcher05} \\
31 & NGC7713     & 23:36:14.99 & $-$37:56:17.1 & 10.29 & 6.7    & 9.90  & 2.17 & 6.06 & 1.4  & 42.4 & \citep{Georgiev16} \\
32 & UGC08041    & 12:55:12.65 & +00:06:59.9   & 17.14 & 6.9    & 8.83  & --   & 6.74 & 10.3 & 34.4 & \citep{Georgiev16,Baldassare22} \\
33 & UGC08516    & 13:31:52.59 & +20:00:04.6   & 14.50 & 5.9    & 8.95  & 1.29 & 6.37 & 11.2 & 34.1 & \citep{Georgiev16} \\
34 & UGC09215    & 14:23:27.13 & +01:43:34.4   & 19.96 & 6.3    & 9.77  & 2.49 & 7.24 & 0.8  & 41.3 & \citep{Georgiev16} \\
35 & VCC0033     & 12:11:07.76 & +14:16:29.3   & 15.00 & $-$4.0 & 9.62  & 0.72 & 6.86 & 7.2  & 38.9 & \citep{Leigh2012,Cote06} \\
36 & VCC0140     & 12:15:12.56 & +14:25:58.4   & 17.15 & $-$2.2 & 9.79  & 2.10 & 7.09 & 7.8  & 48.7 & \citep{Leigh2012,Cote06} \\
37 & VCC0437     & 12:20:48.82 & +17:29:13.4   & 16.90 & $-$3.2 & 9.90  & 2.96 & 6.94 & 61.8 & 70.7 & \citep{Leigh2012,Cote06} \\
38 & VCC0543     & 12:22:19.53 & +14:45:38.8   & 16.04 & $-$2.7 & 9.74  & 2.19 & 6.95 & 10.7 & 19.2 & \citep{Leigh2012,Cote06,Graham19a,Pechetti20,Peng2008,Ferrarese2006} \\
39 & VCC0698     & 12:24:05.02 & +11:13:05.0   & 18.71 & $-$2.0 & 9.98  & 1.45 & 6.41 & --   & 62.0 & \citep{Graham19a,Peng2008,Boker02,Ferrarese2006} \\
40 & VCC0751     & 12:24:48.36 & +18:11:42.4   & 15.59 & $-$2.7 & 9.63  & 1.79 & 6.98 & 22.0 & 65.0 & \citep{Leigh2012,Cote06} \\
41 & VCC0856     & 12:25:57.93 & +10:03:13.6   & 16.83 & $-$3.3 & 9.35  & 1.22 & 5.93 & --   & 33.0 & \citep{Graham19a,Peng2008,Ferrarese2006,Forbes11} \\
42 & VCC0939     & 12:26:47.23 & +08:53:04.6   & 16.83 & 6.3    & 9.48  & 4.30 & 6.55 & 4.8  & 45.8 & \citep{Georgiev16,Baldassare22} \\
43 & VCC1087     & 12:28:14.88 & +11:47:23.4   & 16.67 & $-$4.1 & 9.51  & 1.53 & 6.03 & --   & 45.0 & \citep{Graham19a,Spavone2020,Boker02,Peng2008,Ferrarese2006} \\
44 & VCC1125     & 12:28:43.31 & +11:45:18.1   & 16.04 & $-$1.9 & 9.93  & 7.15 & 7.32 & 38.1 & 65.3 & \citep{Leigh2012,Cote06} \\
45 & VCC1192     & 12:29:30.25 & +07:59:34.3   & 16.50 & $-$4.8 & 9.27  & 0.60 & 5.79 & --   & 58.2 & \citep{Graham19a,Peng2008,Wegner2003,Ferrarese2006} \\
46 & VCC1250     & 12:29:59.08 & +12:20:55.2   & 17.62 & $-$2.9 & 9.57  & 1.35 & 6.31 & --   & 69.0 & \citep{Graham19a,Peng2008,Boker02,Ferrarese2006} \\
47 & VCC1261     & 12:30:10.33 & +10:46:46.1   & 18.91 & $-$4.8 & 9.69  & 1.71 & 6.23 & --   & 50.0 & \citep{Graham19a,Pechetti20,Peng2008} \\
48 & VCC1355     & 12:31:20.19 & +14:06:54.7   & 16.90 & $-$5.0 & 9.75  & 1.77 & 6.80 & 23.8 & 64.8 & \citep{Leigh2012,Cote06} \\
49 & VCC1422     & 12:32:14.21 & +10:15:05.3   & 15.35 & $-$4.8 & 9.58  & 1.82 & 6.13 & --   & 36.0 & \citep{Graham19a,Boker02,Peng2008}         \\
50 & VCC1431     & 12:32:23.39 & +11:15:46.7   & 17.50 & $-$4.8 & 9.97  & 2.43 & 7.11 & --   & 55.0 & \citep{Leigh2012,Cote06}            \\
51 & VCC1488     & 12:33:13.44 & +09:23:50.5   & 18.30 & $-$3.5 & 9.62  & 0.70 & 6.83 & 1.3  & 18.6 & \citep{Leigh2012,Cote06}            \\
52 & VCC1489     & 12:33:13.90 & +10:55:42.7   & 18.82 & $-$5.0 & 9.68  & 0.34 & 6.66 & 5.6  & 25.5 & \citep{Leigh2012,Cote06}            \\
53 & VCC1528     & 12:33:51.62 & +13:19:20.8   & 15.41 & $-$4.4 & 9.88  & 1.84 & 7.05 & 7.2  & 51.9 & \citep{Leigh2012,Cote06}            \\
54 & VCC1539     & 12:34:06.74 & +12:44:29.8   & 16.35 & $-$5.0 & 9.85  & 0.69 & 6.66 & 32.1 & 28.5 & \citep{Leigh2012,Cote06}            \\
55 & VCC1545     & 12:34:11.53 & +12:02:56.3   & 17.60 & $-$5.0 & 9.94  & 1.33 & 6.96 & 11.5 & 45.6 & \citep{Leigh2012,Cote06}            \\
56 & VCC1661     & 12:36:24.78 & +10:23:04.8   & 17.76 & $-$5.0 & 9.67  & 1.08 & 6.63 & 51.4 & 64.9 & \citep{Leigh2012,Cote06}            \\
57 & VCC1695     & 12:36:54.85 & +12:31:12.3   & 16.50 & $-$3.0 & 9.76  & 2.05 & 6.85 & 10.6 & 54.5 & \citep{Leigh2012,Cote06}            \\
58 & VCC1861     & 12:40:58.56 & +11:11:04.2   & 15.23 & $-$4.8 & 9.87  & 2.54 & 6.99 & 60.8 & 58.6 & \citep{Leigh2012,Cote06}            \\
59 & VCC1871     & 12:41:15.73 & +11:23:14.1   & 15.49 & $-$5.0 & 9.35  & 0.52 & 5.76 & --   & 51.0 & \citep{Graham19a,Boker02,Peng2008,Ferrarese2006}       \\
60 & VCC1883     & 12:41:32.75 & +07:18:53.6   & 16.60 & $-$2.0 & 10.22 & 2.01 & 6.59 & --   & 60.8 & \citep{Graham19a,Peng2008,Ferrarese2006,Wegner2003}       \\
61 & VCC1895     & 12:41:51.98 & +09:24:10.4   & 15.80 & $-$3.0 & 9.63  & 0.87 & 6.87 & 2.0  & 25.6 & \citep{Leigh2012,Cote06}            \\
62 & VCC2019     & 12:45:20.42 & +13:41:33.6   & 17.06 & $-$3.9 & 9.00  & 1.27 & 5.86 & --   & 37.0 & \citep{Graham19a,Pechetti20,Peng2008,Ferrarese2006}       \\
63 & VCC2048     & 12:47:15.30 & +10:12:12.9   & 16.60 & $-$4.5 & 9.73  & 2.77 & 7.08 & 21.1 & 67.5 & \citep{Leigh2012,Cote06}            \\
64 & VCC2050     & 12:47:20.64 & +12:09:59.1   & 17.01 & $-$4.0 & 9.63  & 0.78 & 6.83 & 6.6  & 26.0 & \citep{Leigh2012,Cote06}            \\
65 & FCC100      & 03:31:47.58 & $-$35:03:06.7 & 19.48 & $-$2.2 & 8.70  & 1.10 & 6.40 & --   & 35.0 & \citep{su22,Rijcke05,Eftekhari2022}          \\
66 & FCC106      & 03:32:47.65 & $-$34:14:19.3 & 20.00 & $-$2.8 & 8.90  & 1.00 & 6.70 & --   & 40.0 & \citep{su22,Turner2012,Gomez23}          \\
67 & FCC119      & 03:33:33.84 & $-$33:34:23.9 & 20.00 & $-$2.9 & 9.00  & 1.70 & 6.81 & 9.7  & 20.0 & \citep{Fahrion2021}               \\
68 & FCC135      & 03:34:30.86 & $-$34:17:51.0 & 18.73 & $-$3.0 & 8.70  & 1.41 & 6.30 & --   & 24.0 & \citep{su22,Briceno2018,Eftekhari2022}          \\
69 & FCC136      & 03:34:29.48 & $-$35:32:47.0 & 18.80 & $-$2.4 & 9.10  & 1.29 & 6.60 & --   & 64.3 & \citep{su22,Rijcke05}             \\
70 & FCC148      & 03:35:16.82 & $-$35:15:56.4 & 20.00 & $-$2.2 & 9.76  & 2.70 & 8.37 & 22.0 & 56.0 & \citep{Fahrion2021,Lyubenova2019,Vanderbeke2011}         \\
\hline
\end{tabular}
\label{tab:sample-continue}
\end{adjustwidth}
\end{table*}
\begin{table*}
\footnotesize
\begin{adjustwidth}{-\extralength}{0cm}
    \caption*{{\textbf{Table 3.}  Continued.}}
    \centering\vspace{-2mm}
\begin{tabular}{cccccccccccc}
\hline\hline
No.&Object&RA&Decl.&$D$&Hubble&$M_{\star, \rm gal}$&$R_{\rm e,gal}$&$M_{\rm NSC}$&$R_{\rm e,NSC}$& $\sigma_{\star,\rm NSC}$& Ref. \\
 & name & (h:m:s) & (d:m:s) &(Mpc)&type&(\Msun)&(kpc)&(\Msun)&(pc)&(\kms)& ~ \\ 
(1) & (2) & (3) & (4) & (5) & (6) & (7) & (8) & (9) & (10) & (11) & (12)\\\hline
71 & FCC150      & 03:35:24.09 & $-$36:21:49.6 & 18.30 & $-$4.3 & 8.70  & 0.50 & 6.30 & --   & 63.8 & \citep{su22,Rijcke05}             \\
72 & FCC176      & 03:36:45.25 & $-$36:15:22.4 & 16.67 & 0.1    & 9.70  & --   & 7.30 & --   & 56.4 & \citep{su22,Wegner2003}           \\
73 & FCC177      & 03:36:47.49 & $-$34:44:22.6 & 20.00 & $-$1.9 & 9.93  & 3.50 & 7.83 & 9.0  & 59.9 & \citep{Fahrion2021,Lyubenova2019,Vanderbeke2011} \\
74 & FCC182      & 03:36:54.30 & $-$35:22:28.8 & 19.40 & $-$2.8 & 9.18  & 1.00 & 6.03 & 3.6  & 60.1 & \citep{Fahrion2021,Wegner2003,Turner2012}        \\
75 & FCC202      & 03:38:06.54 & $-$35:26:24.4 & 19.90 & $-$3.1 & 9.03  & 1.00 & 6.76 & 4.5  & 64.9 & \citep{Fahrion2021,Wegner2003,Turner2012}        \\
76 & FCC207      & 03:38:19.27 & $-$35:07:44.7 & 19.10 & $-$4.5 & 8.50  & 0.78 & 6.00 & --   & 60.9 & \citep{su22,Rijcke05}             \\
77 & FCC222      & 03:39:13.30 & $-$35:22:17.2 & 20.00 & $-$2.2 & 8.80  & 1.40 & 6.45 & --   & 45.0 & \citep{Fahrion2021}               \\
78 & FCC223      & 03:39:19.55 & $-$35:43:35.0 & 20.00 & $-$4.9 & 8.78  & 1.60 & 6.38 & --   & 17.0 & \citep{Fahrion2021,Eftekhari2022} \\
79 & FCC245      & 03:40:33.86 & $-$35:01:21.4 & 20.00 & $-$4.3 & 8.77  & 1.30 & 6.05 & --   & 39.0 & \citep{Fahrion2021}               \\
80 & FCC255      & 03:41:03.61 & $-$33:46:45.5 & 20.00 & $-$2.1 & 9.70  & 1.30 & 6.98 & --   & 49.0 & \citep{Fahrion22,Vanderbeke2011}  \\
81 & FCC288      & 03:43:22.63 & $-$33:56:25.1 & 17.66 & $-$2.2 & 8.70  & 0.82 & 6.20 & --   & 48.5 & \citep{su22,Rijcke05}             \\
82 & FCC301      & 03:45:03.56 & $-$35:58:21.4 & 20.00 & $-$3.3 & 9.30  & 1.10 & 6.91 & --   & 54.0 & \citep{Fahrion2021,Vanderbeke2011} \\
83 & FCC303      & 03:45:14.09 & $-$36:56:12.3 & 19.50 & $-$4.3 & 8.70  & 1.10 & 7.00 & --   & 27.0 & \citep{su22,Rijcke05,Briceno2018} \\
84 & FCC310      & 03:46:13.74 & $-$36:41:46.8 & 20.00 & $-$1.9 & 9.73  & 3.50 & 7.81 & 31.0 & 60.4 & \citep{Fahrion2021,Lyubenova2019} \\
85 & FCC335      & 12:21:40.55 & +11:29:59.7   & 16.83 & 8.5    & 9.30  & 1.62 & 7.00 & 0.9  & 74.1 & \citep{Georgiev16,Baldassare22}   \\
\hline
\end{tabular}
\parbox[t]{1.0\textwidth}{\textit{Notes:} Columns 2--4: The galaxies name and their J2000 coordinates. Columns 5--8: The distance derived from NED database, galaxy morphology, galaxy stellar mass, and galaxy effective radius. For galaxies in the Fornax cluster lacking individual distance measurements, a common cluster distance of $D_{\rm Fornax} = 20.00$~Mpc has been adopted. Columns 9--11:  NSC mass, NSC effective radius, and stellar velocity dispersion of the NSC. Column 12: references.}
\end{adjustwidth}
\end{table*}

\subsection{Sample properties}\label{sec:sample-properties}

\begin{figure}
    \centering
    \includegraphics[width=0.5\columnwidth]{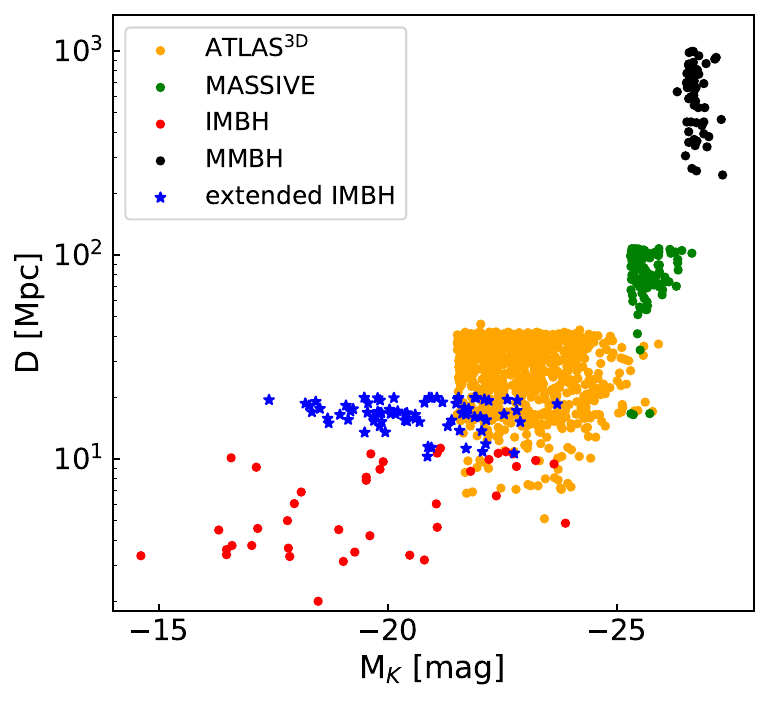}
    \caption{The distribution of our extended-HARMONI IMBH sample on distance vs. $K$s-band absolute magnitude comparing with the previous surveys: HARMONI MMBH \citep{Nguyen23}, MASSIVE \citep{Ma14}, ATLAS$^{\rm 3D}$ \citep{Cappellari11} and HARMONI IMBH \citep{Nguyen2025b}.} 
    \label{fig:mk-distance}
\end{figure}

We summarized in Table \ref{tab:selected_sample} our expanded HARMONI IMBH sample, spanning a distance range of $10 \lesssim D \lesssim 20$ Mpc and an absolute $K_s$-band magnitude range of $-23.7 \lesssim M_K/{\rm mag} \lesssim -17.4$. Figure \ref{fig:mk-distance} presents  our selected targets on the distance–magnitude plane and compare them with previous galaxy surveys, including MMBH \citep{Nguyen23}, MASSIVE \citep{Ma14}, ATLAS$^{\rm 3D}$ \citep{Cappellari11}, and the HARMONI IMBH survey \citep{Nguyen2025b}. It overlaps with the ATLAS$^{\rm 3D}$ survey by 2.2 mag ($M_K < -21.5$ mag) and focuses more on lower-mass galaxies. Additionally, our sample is distinct from the 10 Mpc HARMONI IMBH sample \citep{Nguyen2025b}, although both samples cover nearly the same stellar mass range of $10^8$ to $5\times 10^{10}$ \Msun. Our selected galaxies also exhibit central stellar velocity dispersions ranging from 16 to 75 \kms.

\begin{figure}[H]
\centering
    \includegraphics[width=0.85\textwidth]{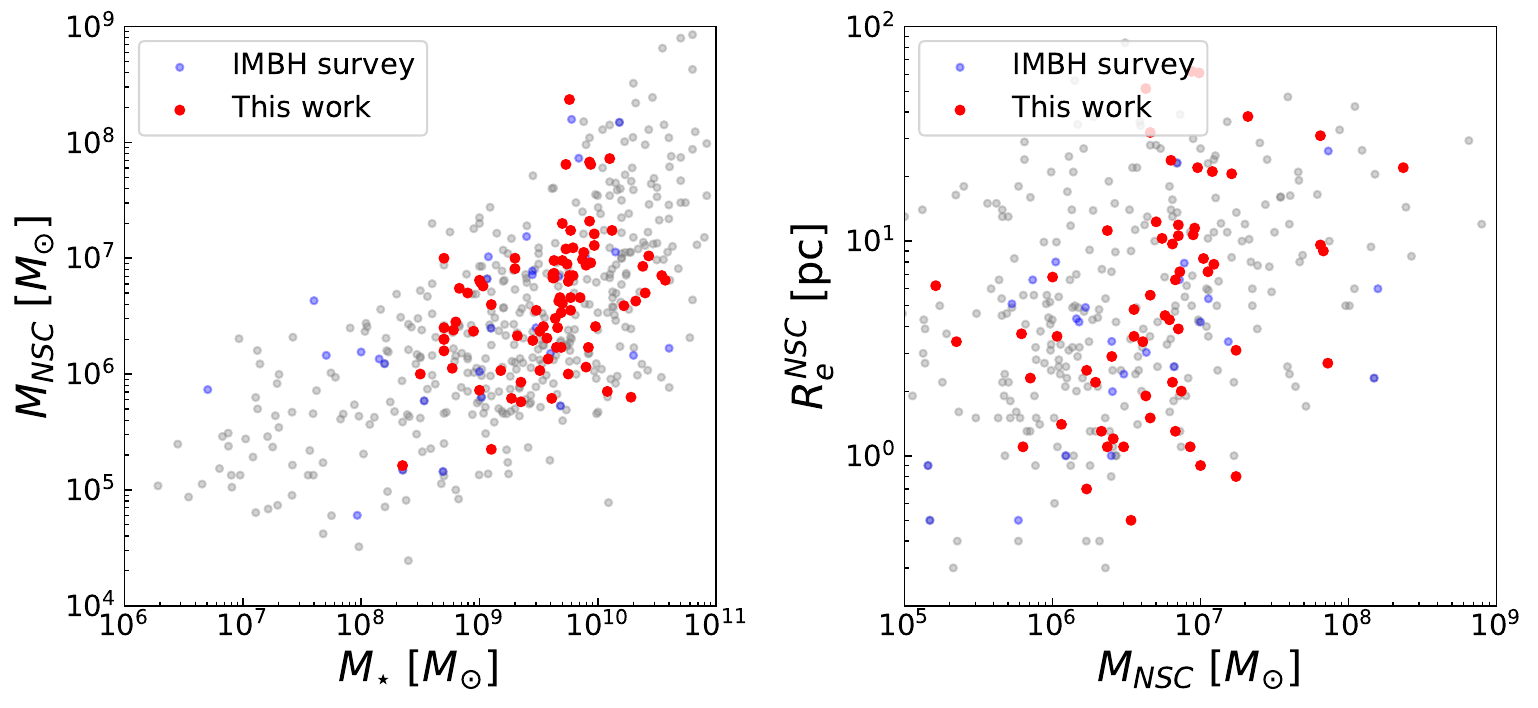}
    \caption{{\bf Left:} The $M_{\star}$--$M_{\rm NSC}$ relation shows the correlation between galaxy mass and NSC mass across various Hubble types \citep{Erwin12, Georgiev16, Spengler17, Ordenes-Briceno18b, Sanchez-Janssen19}. {\bf Right:} The $M_{\star}$--$R_e$ relation illustrates the connection between galaxy mass and the NSC's effective radius \citep{Cote06, Georgiev14, Norris14, Baldassare22}. For comparison, we included the 10 Mpc HARMONI IMBH sample \citep{Nguyen2025b}.}
    \label{NSCs}
\end{figure}

This expanded HARMONI IMBH candidates comprises 85 galaxies, including 29 and 20 members of the Virgo (prefix VCC) and Fornax (prefix FCC) Clusters, respectively, along with 36 isolated dwarf galaxies or members of other groups, as detailed in Table \ref{tab:sample-continue}. It spans a diverse range of galaxy morphologies, consisting of 27\% ellipticals, 29\% lenticulars, 39\% spirals, and 5\% irregulars. These galaxies host luminous NSCs with masses in the range $1.6 \times 10^5 < M_{\rm NSC} < 2.3 \times 10^8$~\Msun\ and effective radii spanning $0.5 < R_{\rm e} < 62$ pc, as shown in Figure \ref{NSCs} for the $M_{\star}$–$M_{\rm NSC}$ and $M_{\star}$–$R_{\rm e}$ relationships.

A key characteristic of our sample is its predominance of more massive NSCs, which are larger in both mass and size compared to those in the previous HARMONI IMBH candidates \citep{Nguyen2025b}. This distinction arises because our survey extends to greater distances. Despite our thorough efforts to select the most suitable targets from all available databases and literature, our sample represents only a small subset of the total potential candidates. A comprehensive, high-resolution, and deep imaging survey remains necessary to construct a complete census of NSCs and uncover the elusive IMBH population in nearby nucleated dwarf galaxies.

\subsection{Searching tip-tilt stars and observing strategies} \label{sec:ngs} 

The AO system is crucial for achieving high spatial resolution observations with the ELT. In LTAO mode, the system is supported by laser guide stars (LGS) and operates with a natural guide star (NGS) that can be up to 10,000 times fainter than those required by the Gemini and VLT AO systems. The NGS serves as a reference star to correct tip/tilt distortions and must have an $H$-band magnitude of $m_H < 20.4$ ABmag \citep{Thatte16}.

Since the ELT will reach the lowest sensitivity limits for NGSs, current NGS search tools are inadequate. To identify suitable NGSs, we utilized the Gaia Data Release 3\footnote{\url{https://www.cosmos.esa.int/web/gaia/dr3}}, which provides an all-sky catalog of $\approx$$10^9$ point sources with a magnitude limit of $m_V \lesssim 20$ Vegamag \citep{Pancino16}. We employed the \textsc{cone\_search\_async} function from the {\tt astroquery} package\footnote{\url{https://github.com/astropy/astroquery}} \citep{Ginsburg19} to retrieve stellar coordinates and photometric information, which will then converted to the $H$-band ABmag using the following relation \citep{Busso22}:
\begin{equation} 
G - H = a + b\left(G_{BP} - G_{RP} \right) + c \left(G_{BP} - G_{RP} \right)^2,
\end{equation}
Here, the $G$-band magnitude and the integrated blue-to-red passband color ($G_{\rm BP} - G_{\rm RP}$) were retrieved from Gaia, while the coefficients $a$, $b$, and $c$ were taken from Table 5.9 of \citep{Busso22}.

\begin{figure}
\begin{adjustwidth}{-\extralength}{0cm}
\centering
\includegraphics[width=1.2\textwidth]{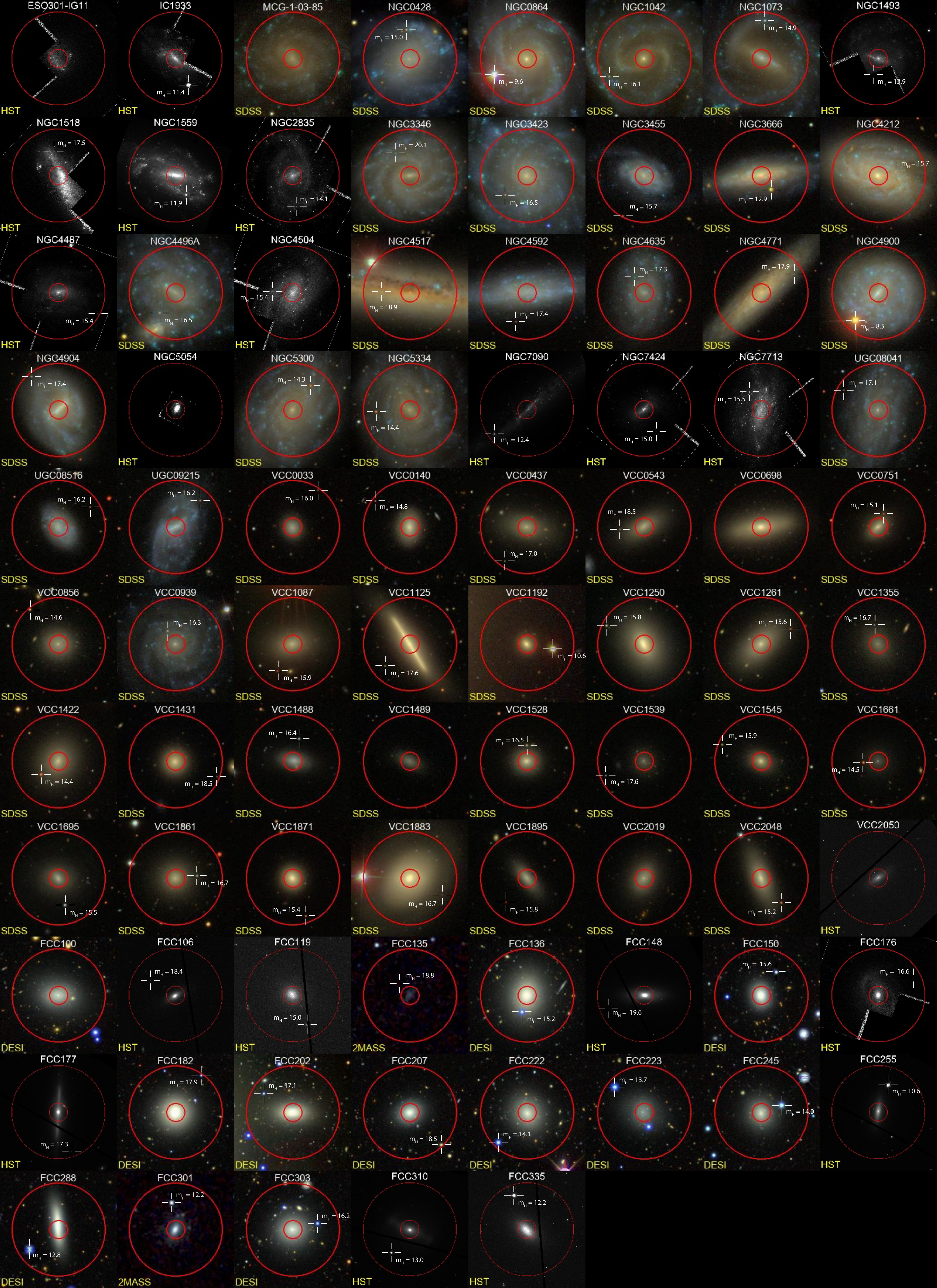}
\caption{Images from the HST, the Dark Energy Spectroscopic Instrument (DESI) Legacy Surveys DR10,  The Two Micron All-Sky Survey (2MASS), and Sloan Digital Sky Survey (SDSS) of our extended HARMONI IMBH sample. Two red circles indicate radii of 12\arcsec\ (inner) and 60\arcsec\ (outer) from the galaxy center, marking the optimal region for selecting NGSs for AO correction. These stars are indicated by white crosses, with their respective $H$-band magnitudes estimated from Gaia.} 
\label{fig:sample}
\end{adjustwidth}
\end{figure}

For each galaxy, we selected one star based on the following criteria: (1) it lies within an annulus of $12\arcsec < r < 60\arcsec$ from the galaxy center, and (2) it has an $H$-band magnitude of $m_H < 20.4$ ABmag. The selected stars are marked with white crosses, with their $H$-band magnitudes labeled next to them in Figure \ref{fig:sample}.

\begin{figure}
    \centering
    \includegraphics[width=0.95\textwidth]{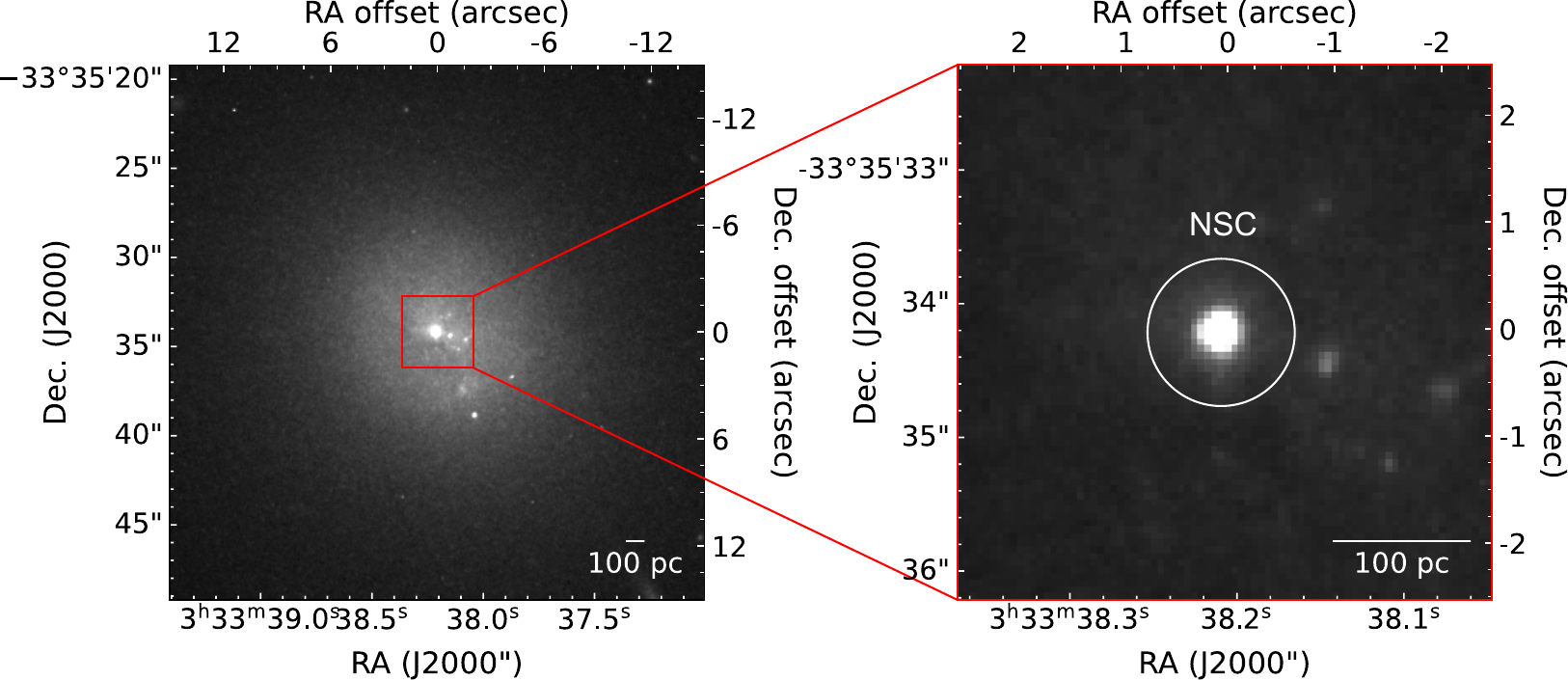}
    \caption{\textbf{Left:} The HST/ACS WFC F850LP image of FCC 119 ($15\arcsec\times15\arcsec$ or $1.5\times1.5$~kpc$^2$) shows its large-scale morphology.  \textbf{Right:} A zoom-in to the central region ($5\arcsec \times 5\arcsec$) highlights its NSC.}
    \label{fig:enter-label}
\end{figure}

\subsection{The simulated target: FCC 119}\label{Sec:fcc119} 

We aim to determine the required exposure time (sensitivity) for observing faint dwarf galaxies with ELT/HARMONI, ensuring to resolve the stellar kinematic signature of central IMBHs and accurately measure their masses dynamically. At $\approx$20~Mpc, galaxies in the Fornax Cluster appear smaller in angular size and fainter in apparent surface brightness compared to those in the Virgo Cluster ($D \approx 16.5$~Mpc), making difficulty in measuring their IMBH. 

To assess this, we selected FCC~119 ($m_B = 15.44$ mag), the second faintest galaxy in the Fornax Cluster \citep{Turner2012}, which is fainter than nearly all Virgo Cluster galaxies in our sample \citep{Cote06}. FCC 119 is an S0 dwarf galaxy located at $03^{\rm h}33^{\rm m}33\fs7$–$33\degr34\arcmin18\arcsec$. Its stellar mass is estimated to be \Mstar\ $\approx (1.06-1.4)\times10^9$ \Msun, with an effective radius of $R_{\rm e} = 1.33$ kpc \citep{Turner2012, Liu19, Spavone2022}, based on $r$-band observations from OmegaCAM at the ESO/VLT Survey Telescope \citep[VST;][]{Kuijken11, Schipani2012}. The galaxy has a SB of 20.1 mag arcsec$^{-2}$ at a radius of 1\arcsec\ in the $z$-band \citep{Turner2012}. FCC~119 hosts a bright, young NSC with a mass of $M_{\rm NSC} \approx 6.5 \times 10^6~{\rm M_\odot}$ \citep{Fahrion2021}, identified in \textit{HST}/Advanced Camera for Surveys (ACS) imaging (Figure~\ref{fig:enter-label}; \citep{Jordan2007, Turner2012}). The NSC has an estimated age of $1.7 \pm 0.5$~Gyr and a metallicity of $[{\rm Fe/H}] = -1.02 \pm 0.02$~dex \citep{Fahrion2021}, whereas its surrounding regions are significantly older, with stellar ages increasing from about 5 to 10~Gyr \citep{Iodice2019a, Fahrion2021}.

The galaxy's kinematic inclination was determined from optical data obtained with the VLT/Multi Unit Spectroscopic Explorer (MUSE), yielding $i = 64^\circ.9 \pm 5^\circ.4$. Stellar kinematic analysis of these data revealed rotation with $V \approx 10$ \kms\ at $r\approx2\arcsec$, and a velocity dispersion of $\sigma_\star \approx 20$ \kms\ \citep{Ding2023, Iodice2019a}. The presence of emission lines exclusively within the NSC suggests ongoing star formation concentrated in the nuclear region.

\subsection{Fornax \& Virgo Clusters}\label{Sec:fornaxfcc119}

\subsubsection{Virgo Cluster}\label{Sec:virgo}

Although the Virgo Cluster is located in the Northern Hemisphere, some targets can still be observed with the ELT (Table \ref{tab:parent_criteria}). As the most extensively studied cluster to date, Virgo contains $\approx 2000$ members \citep{Binggeli85} at a distance of $\approx 16.5$ Mpc \citep{Mei07}. It is the most massive galaxy cluster in the Local Supercluster, with an estimated virial mass of $M_{\rm vir} \approx (1.4 - 6.3) \times 10^{14}~\Msun$ \citep{Urban11,McLaughlin99, Kashibadze20}, of which $\sim 93\%$ is found to be non-baryonic \citep{McLaughlin99}. This cluster exhibits a strong segregation of Hubble types: its center is dominated by early-type galaxies (ETGs), while dwarf elliptical/dwarf S0 (dE/dS0) galaxies dominate the overall census, comprising $\approx$75\% of the members \citep{Binggeli87, Ferguson89}. Several large surveys have attempted to study the Virgo Cluster in multiple wavelengths, including X-ray \citep{Truemper93, Bohringer94, Gallo08}, optical \citep{Cote06, Ferrarese12}, near-infrared \citep{Skrutskie06, Lawrence07, McDonald09}, far-infrared \citep{Davies10HeViCS, Boselli10}, and radio \citep{Becker95, White97, Wilson09}. These surveys have provided the groundwork for investigating dwarf galaxies and predicting the presence of IMBHs \citep{Volonteri08a, Gallo10, Graham19a, Graham19b}.

\subsubsection{Fornax Cluster}\label{Sec:fornax}

Fornax is an excellent galaxy cluster for studying galaxy evolution in the Southern Hemisphere. With a virial mass of $M_{\rm vir} = 7 \times 10^{13}$ \Msun\ \citep{Drinkwater2001}, it is the second largest cluster on the sky after Virgo, which contains numerous ETGs at a distance of 20 Mpc \citep{Blakeslee2009}. This cluster has been investigated in multi-wavelength surveys: X-ray \citep{Su2017}, optical \citep{Kuijken02, Scharf2005, Jordan2007, Munoz2015, Iodice2016}, NIR \citep{Davies2013}, and radio \citep{Loni2021}. The distance to the cluster currently prevents dynamical measurements of its dwarf members’ \Mbh\ because their $r_{\rm SOI}$ is far below the resolving power of current telescopes, except for FCC 213 \citep{Houghton2006, Gebhardt2007} and FCC~47 \citep{Thater23}. However, given the significant number of its dwarf ETGs and GCs, Fornax represents a promising location for searching for IMBHs in the local volume with the ELT.

\section{Stellar mass model of FCC 119}\label{Sec:Stellar-mass-model}

\subsection{\textit{HST} observations}\label{Sec:hst}

We used Hubble Space Telescope (HST) Advanced Camera for Surveys (ACS) Wide Field Channel (WFC) F850LP observations of the Fornax Cluster (PI: Andres Jordan) from the Hubble Legacy Archive (HLA) to constrain the radial SB profile and stellar mass model for FCC 119. The image has a pixel scale of $0\farcs05$ and a total exposure time of 1220 seconds, composed of three frames. 

\begin{figure}
    \centering
    \includegraphics[width=0.7\columnwidth]{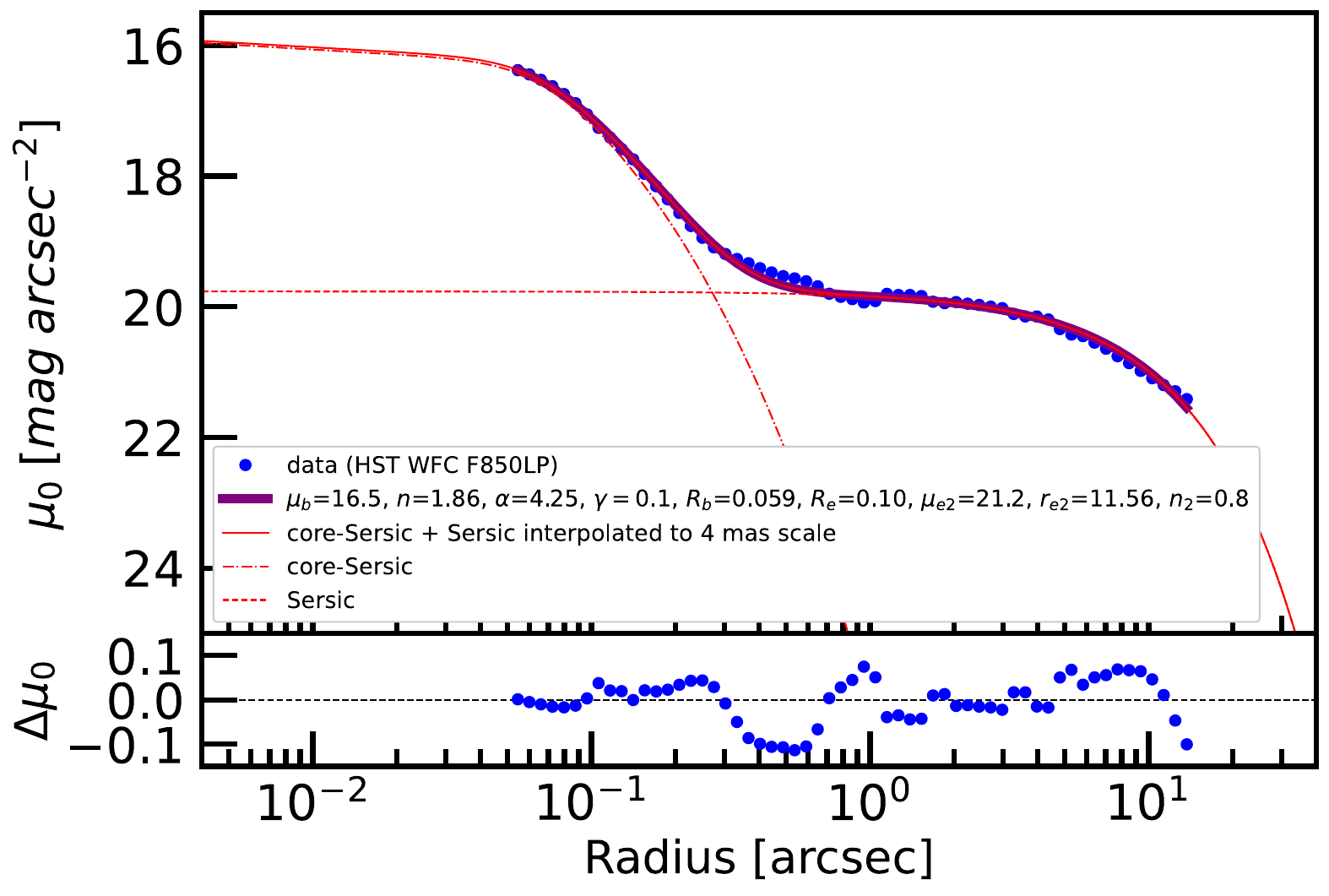}
    \caption{{\bf Top:} The HST ACS/WFC F850LP radial SB profile of FCC 119 (blue dots) is shown simultaneously with the best-fit model (purple-thick line), which is a combination of a core-S\'ersic (red-dotted line) and a S\'ersic (red-dashed line) function, extrapolated into both the unresolved region of $0\farcs004$ and the extended region beyond 20$\arcsec$. The best-fit parameters of the best-fit model are shown in the legend. {\bf Bottom:} The residual ({\tt data-model}) between the SB and the best-fit combined function.}
    \label{fig:IRAFextrapolate}
\end{figure}
\subsection{HST point spread function (PSF) simulations}\label{Sec:psf}

We generated the F850LP PSF using the {\tt Tiny Tim} package \citep{Krist11} to accurately derive the intrinsic SB of FCC 119. This software constructs model PSFs tailored to the specific instrument, detector chip, chip position, and filter used in the observations. To ensure consistency with the processing of the HST images, we created individual model PSF for each exposure frame (i.e., three PSFs for three exposures), incorporating sub-pixel offsets following a four-point box dither pattern. Additionally, each model PSF was convolved with an appropriate charge diffusion kernel to account for electron leakage into adjacent CCD pixels. The three individual PSFs were then combined and resampled onto a final grid with a pixel size of $0\farcs05$ using {\tt Drizzlepac}/{\tt AstroDrizzle} \citep{Avila12}, producing the final PSF, which has a FWHM of $0\farcs11$.

\subsection{F850LP surface brightness profile}\label{Sec:sb}

To mitigate contamination from foreground stars, we first masked them in the HST image using {\tt SExtractor}\footnote{{\tt SExtractor}: \url{https://github.com/astromatic/sextractor/}} \citep{Bertin1996}, a tool designed to extract light sources from astronomical images and classify them as galaxies or point sources (stars) based on the ``stellarity index'' in the {\tt CLASS\_STAR} parameter. We set a seeing FWHM of $0\farcs2$ and a detection threshold ({\tt Detect\_thresh}) of 20 mag. Objects with a stellarity index greater than 0.5 were classified as stars and subsequently masked from the image. 

Second, we extracted the radial SB profile along the galaxy’s semi-major axis from the masked image using the {\tt ellipse} task with the {\tt Image Reduction and Analysis Facility (IRAF)} \citep{Jedrzejewski87}. This task measures the integrated flux (in counts s$^{-1}$) within concentric annular rings while accounting for variations in position angle and ellipticity. A photometric zero-point of 24.87 mag was applied, calculated using the {\tt ACS Zeropoints Calculator} from the {\tt acstools} package\footnote{{\tt acstools} v3.7.0: \url{https://pypi.org/project/acstools/}}. Additionally, a solar absolute magnitude of 4.5 mag \citep{Willmer2018} and a Galactic extinction value of 0.017 mag \citep{Schlafly11} were adopted. During this process, we deconvolved the HST image using the final {\tt Tiny Tim} PSF (Section \ref{Sec:psf}). The integrated flux within each annulus was then converted to mag arcsec$^{-2}$.

Next, we fitted the radial SB profile using a combination of a core-Sérsic \citep{Graham03, Trujillo04} and a Sérsic \citep{Sersic68} function, which describe the light profiles of the NSC and the disk component, respectively. The core-Sérsic function has the following analytical form:
\begin{equation}
	I(r) = I' \left[1 + \left(\frac{R_b}{r}\right)^\alpha\right]^{\gamma/\alpha} 
	\exp\left[{-b_{n_{\rm NSC}} \left(\frac{r^\alpha + R_b^\alpha}{R_e^\alpha}\right)^{1/(\alpha\, n_{\rm NSC})}}\right] 
    \label{eq:core-sersic}
\end{equation} 
where $I'=I_b 2^{-\gamma/\alpha}\exp\left[b_{n_{\rm NSC}}2^{1/(\alpha\, n_{\rm NSC})}\left( \dfrac{R_b}{R_e}\right)^{1/n_{\rm NSC}} \right]$.
Here, $R_b$ is the break radius, marking the point at which the radial SB profile transition from the outer S\'ersic to the inner power law regimes, while the sharpness of this transition is controlled by $\alpha$. $I_b$ is the intensity at $R_b$, which will be converted into SB values ($\mu_b$). The shape of the outer S\'ersic part ($r>R_b$) is determined by its S\'ersic index ($n_{\rm NSC}$) and effective radius ($R_e$), while the inner-power law part ($r<R_b$) is sharped by the power-law index $\gamma$. And the constant $b_{n_{\rm NSC}}$ is approximated as $b_{n_{\rm NSC}} = 2n_{\rm NSC} - \dfrac{1}{3} + \dfrac{4}{405n_{\rm NSC}} + \dfrac{46}{25515n^2_{\rm NSC}} + {\rm O}\left(\dfrac{1}{n^3_{\rm NSC}}\right)$ \citep{Ciotti99}.

Given that the observable scale of the HST image is 5 times coarser than that of HARMONI, detailed stellar information near the IMBH's SOI is unavailable. To address this, \citep{Nguyen2025b} assumed a core-Sérsic function with a power-law index of $\gamma = 0.1$ for the NSC. We also adopt this assumption in this work.

The Sérsic function \citep{Sersic68} describes the SB profile of the galaxy’s disk component, characterized by the half-light radius ($R_{e,\rm disk}$), Sérsic index ($n_{\rm disk}$), and intensity ($I_{e,\rm disk}$) at $R_{e,\rm disk}$, expressed as:
\begin{equation}
    I(r)=\left(I_{e,\rm disk}\right)\exp\left\{-b_{n_{\rm disk}}\left[\left(\dfrac{r}{R_{e,\rm disk}}\right)^{1/n_{\rm disk}}-1\right]\right\}
    \label{eq:sersic}
\end{equation}
where, $b_{n_{\rm disk}}$ is determined so that $\gamma(2n_{\rm disk};b_{n_{\rm disk}}) = \dfrac{1}{2}\Gamma(2n_{\rm disk})$, where $\Gamma$ and $\gamma$ respectively refer to the Gamma and lower incomplete Gamma function \citep{Ciotti1991}. And, we adopt $b_{n_{\rm disk}} \approx 2n_{\rm disk}-1/3$. 

We employed a non-linear least-squares fitting algorithm using the {\tt Python MPFIT}\footnote{\url{http://purl.com/net/mpfit.}} function \citep{Markwardt09} to iteratively fit a combined core-Sérsic and Sérsic function to the intrinsically deconvolved radial SB profile of FCC 119. The best-fit parameters include those for the core-Sérsic function: $n_{\rm NSC} = 1.86 \pm 0.41$, $\alpha = 4.25 \pm 0.35$, $R_{\rm e,NSC} = 0\farcs100 \pm 0\farcs006$, $R_{b} = 0\farcs059 \pm 0\farcs004$, and $\mu_{b} = 16.5 \pm 0.4$ mag arcsec$^{-2}$; and for the Sérsic function: $n_{\rm disk} = 0.8 \pm 0.2$, $R_{\rm e,disk} = 11\farcs56 \pm 3\farcs47$, and $\mu_{\rm e,disk} = 21.2 \pm 0.8$ mag arcsec$^{-2}$. Based on these best-fit parameters, we extrapolated the SB profile toward the galaxy center down to a scale of $0\farcs01$, as required for this extended HARMONI IMBH sample, and present the results in Figure \ref{fig:IRAFextrapolate}.


\subsection{MGE mass model}\label{Sec:mge}

We used the best-fitting parameters from the combined core-Sérsic and Sérsic functions to construct an MGE model using the \texttt{mge.fit\_1d} procedure from the \textsc{MgeFit} package\footnote{v5.0.15: \url{https://pypi.org/project/mgefit/}} \citep{Emsellem94, Cappellari02}. The input logarithmically sampled Sérsic profile was sampled over the range $0.04$–$400R_\mathrm{e}$ and fitted with nine Gaussians to ensure an accurate representation of the galaxy SB distribution. To account for ellipticity, defined as $\epsilon = 1-b/a = 1-q'$, which quantifies the elongation of the galaxy along its major axis, we used the \textsc{find\_galaxy} procedure from the \textsc{MgeFit} package to measure $\epsilon$. FCC 119 shows a marked change in ellipticity at a radius of $r = 6\arcsec$. Beyond this radius, we find $\epsilon(r > 6\arcsec) \approx 0.15$, while inside this radius the morphology is nearly spherical, with $\epsilon(r < 6\arcsec) \approx 0$. This ellipticity information was incorporated into the HST/WFC F850LP MGE model presented in Table \ref{tab:MGEfit} (Column 4) in terms of axis ratio $q'$ for each Gaussian.

\begin{table}
\caption{The HST/WFC F850LP MGE model of FCC 119} 
\centering
\begin{tabular}{cccc} 
\hline\hline
$j$&$\Sigma_{\star,j}/$(\Lsun${\;\rm pc^{-2}})$&$\sigma_j({\rm arcsec})$&$q'_j=b_j/a_j$\\
(1) & (2) & (3) & (4)\\ 
\hline
1 & 9428.31 & 0.060  & 1.00 \\
2 & 2535.54 & 0.131  & 1.00 \\
3 & 175.71  & 0.346  & 1.00 \\
4 & 91.52   & 2.679  & 1.00 \\
5 & 243.69  & 6.034  & 1.00 \\
6 & 187.84  & 9.868  & 0.85 \\
7 & 68.45   & 13.762 & 0.85 \\
8 & 10.19   & 17.689 & 0.85 \\
9 & 0.37    & 21.999 & 0.85 \\
\hline
\end{tabular}
\vspace{2mm}
\parbox[t]{\textwidth}{\textit{Notes:} Columns in order: Gaussian component number, luminosity surface density, dispersion along the major axis, and axial ratios.} 
\label{tab:MGEfit}
\end{table}

Finally, we created the stellar mass model from this MGE model by scaling its surface luminosity (Column 2 of Table \ref{tab:MGEfit}) with a constant \ml. Here, we estimated the \ml$_{\rm F850LP}$ of FCC 119 utilizing the empirical correlation between color and \ml\ from \citep{Sande2015}:
\begin{equation}
    \log_{10}(M/L_z) = a_z(g-z) + b_z
    \label{eq:color-to-ml}
\end{equation}
where $a_z=1.25$ and $b_z=-1.11$ are given in Table 3 of \citep{Sande2015}. We adopted the color value of $g-z = 0.63 \pm 0.12$ mag \citep{Turner2012} to infer $M/L_{\rm F850LP} \approx M/L_z \approx 0.6$ (M$_\odot$/L$_\odot$) for FCC 119. We applied this estimated $M/L_{\rm F850LP}$ for our mass model, which yields an NSC mass of $M_{\rm NSC} = 6.5\times 10^6$ \Msun\ and a total galaxy mass of $M_{\star} = 1.15\times10^9$ \Msun. These values are consistent with the estimates previously discussed in Section \ref{Sec:fcc119} within 8\%. 

In this work, we ignored the variation in \ml\ from stellar populations \citep{Mitzkus17, Nguyen17, Thater17} and the distribution of dark matter halos \citep{Navarro96}.

\section{HARMONI IFS Simulations}\label{sec:HARMONI-simulation}

\subsection{Jeans Anisotropic Model (JAM)}\label{sec:jam}

Our extended HARMONI IMBH sample consists of galaxies with NSCs that may exhibit rotational signatures \citep{Seth08}, making them well-suited for a cylindrical projection. To model their kinematics, we employed JAM, assuming a velocity ellipsoid aligned with cylindrical coordinates ($\sigma_z \neq \sigma_r = \sigma_\theta$) in the meridional plane \citep{Cappellari08}. This was implemented by setting the model keyword {\tt align = `cyl'} in the \textsc{jam\_axi\_proj} procedure within the \textsc{JamPy} package\footnote{{\tt jampy} v7.2.4: \url{https://pypi.org/project/jampy/}} \citep{Cappellari20}.

\subsection{MARCS synthetic library of stellar spectra}\label{sec:marcs}

To simulate the HARMONI IFS, we used the \citep{Maraston11} stellar population synthesis (SPS) spectra, which are based on the Model Atmospheres with a Radiative and Convective Scheme (MARCS) synthetic spectra \citep{Gustafsson08}. The MARCS library provides $\approx$52,000 atmospheric models at a high spectral resolution of $R \approx 20,000$, corresponding to an instrumental broadening of $\sigma_{\rm instr} = 6.4$~\kms. It offers fine spectral sampling with 100,724 flux points ($\Delta\lambda \approx 0.065$ \AA) and covers a broad wavelength range of 0.13--20 \micron, making it well-suited for generating mock IFS data to detect the stellar kinematic signatures of IMBHs.

\begin{table}
\centering
\caption{HSIM simulation via {\tt HSIM}.}  
\begin{tabular}{ccc}
\hline\hline
HARMONI grating & Exposure time ({\tt NDIT}) & Sensitivity ({\tt NDIT})\\
  (1)       &      (2)        & (3)  \\  
\hline
 $H$-high  & 16   & 12\\
 $K$-short & 30   & 24\\   
 $K$-long  & 20   & 16\\   
\hline
\end{tabular}\\
\label{table:hsim-exptime}
\parbox[t]{\textwidth}{\textit{Notes:} The total exposure time (and sensitivity) on source is determined as {\tt DIT}$\times${\tt NDIT} with {\tt DIT} = 900 seconds. These estimated time are the science time on sources without accounting for the target acquisition, overhead, and LTAO setup.} 
\end{table}

\subsection{HARMONI instrument and {\tt HSIM} simulator}\label{sec:HARMONI-HSIM}

HARMONI is the first IFS for the ELT, covering optical and NIR wavelengths (0.46–2.46 \micron) and offering 13 spectral gratings with three different spectral resolving powers: $R \approx 3,300$ (low), $\approx$7,100 (medium), and $\approx$17,400 (high). The instrument provides four observational scales: $4\times4$, $10\times10$, $20\times20$, and $30\times60$ mas$^2$, corresponding to fields of view (FoVs) of $0\farcs86 \times 0\farcs61$, $2\farcs14 \times 1\farcs52$, $4\farcs28 \times 3\farcs04$, and $9\farcs12 \times 6\farcs42$, respectively \citep{Zieleniewski15}. In this study, we conducted our HARMONI IFS simulations using the high-spectral-resolution mode with an intermediate spatial resolution of $10\times10$ mas$^2$ spaxel \citep{Nguyen23, Nguyen2025b}.

The HARMONI IFS cubes were generated using \hsim\ \citep{Zieleniewski15}, which applies observational effects—including instrumental noise, atmospheric conditions, celestial sources with physical properties, realistic detector characteristics, and readout noise—to initially noiseless input data cubes (Section \ref{input-noiseless_cubes}). This produces realistic simulated observations \citep{Nguyen23, Nguyen2025b} and ensures that the resulting data closely resemble actual astronomical observations.

It should be noted that the final performance of HARMONI remains subject to possible descoping in the coming years, which could reduce its spatial resolution and spectral resolution. Such changes in the telescope’s specifications would directly affect the IMBH detection limits. Therefore, the results presented here should be regarded as best-case estimates, based on the ELT’s current planned design.

\begin{figure}
    \centering
    \includegraphics[width=\textwidth]{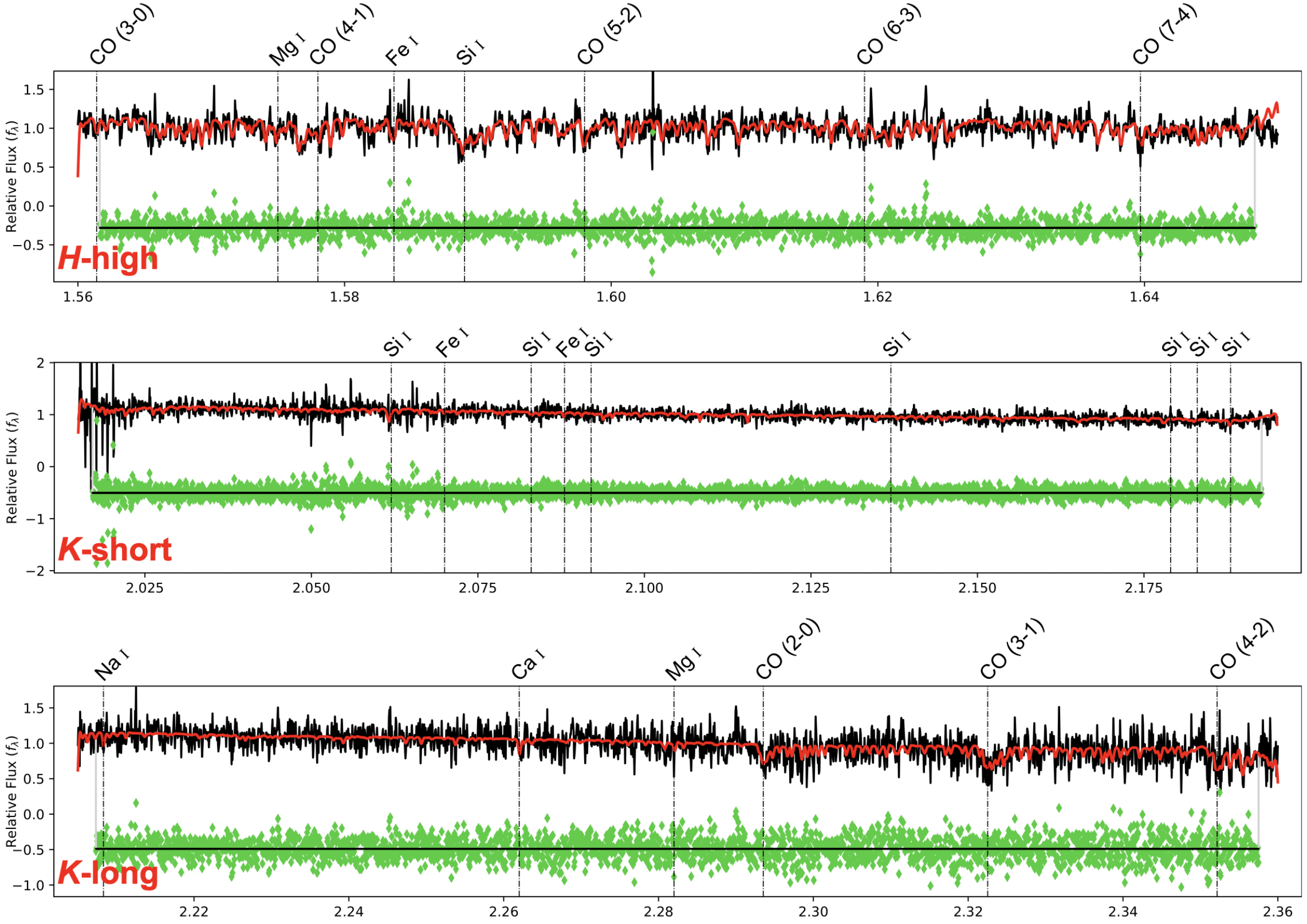}
    \caption{The central {\tt Voronoi}-binned spectra of FCC 119 are presented for the three mock IFS data cubes generated by {\tt HSIM} (black), along with their best-fit {\tt pPXF} models (red). The residuals, calculated as ({\tt data-model}), are shown as green points.}
    \label{fig:ppxf-spectrum}
\end{figure}

\subsection{HARMONI Simulations}\label{sec:createIFS}

We employed the high-spectral-resolution gratings of HARMONI, including the $H$-high (1.538--1.678 \micron), $K$-short (2.017--2.201 \micron), and $K$-long (2.199--2.400 \micron) bands, which include several strong spectral lines that have been widely used in previous stellar kinematic studies \citep{CrespoGomez21, Nguyen2025b}.

The $H$-high band contains numerous metallicity indicators, such as \ion{Mg}{i} ($\lambda$1.487, 1.503, 1.575, 1.711 \micron), \ion{Fe}{i} ($\lambda$1.583 \micron), and \ion{Si}{i} $\lambda$1.589 \micron, along with CO absorption lines, including $^{12}$CO(3-0) $\lambda$1.561 \micron, $^{12}$CO(4-1) $\lambda$1.578 \micron, $^{12}$CO(5-2) $\lambda$1.598 \micron, $^{12}$CO(6-3) $\lambda$1.619 \micron, and $^{12}$CO(7-4) $\lambda$1.640 \micron. In addition, the $K$-long band features prominent CO molecular absorption lines, including $^{12}$CO(2-0) $\lambda$2.293 \micron, $^{12}$CO(3-1) $\lambda$2.322 \micron, $^{12}$CO(4-2) $\lambda$2.351 \micron, and $^{12}$CO(5-3) $\lambda$2.386 \micron, as well as atomic absorption lines, such as \ion{Na}{i} $\lambda$2.207 \micron, \ion{Ca}{i} $\lambda$2.263 \micron, and \ion{Mg}{i} $\lambda$2.282 \micron. Furthermore, the $K$-short grating contains additional atomic lines, including \ion{Si}{i} ($\lambda$2.062, 2.083, 2.092, 2.137, 2.179, 2.183, 2.188 \micron) and \ion{Fe}{i} ($\lambda$2.070, 2.088 \micron). All these spectral features are illustrated in Figure \ref{fig:ppxf-spectrum}.

\subsubsection{Creation of the input-noiseless cubes for {\tt HSIM}}\label{input-noiseless_cubes}

We assumed that the line-of-sight velocity distribution (LOSVD) in the nucleus of FCC 119 is described well by a Gaussian profile. Accordingly, we computed the 2D first ($V$) and second ($V_{\rm rms}$) moments of the stellar component using the \textsc{jam\_axi\_proj} routine from the \textsc{JamPy} package (Section \ref{sec:jam}) \citep{Cappellari20}. The velocity dispersion was then derived as $\sigma_\star=\sqrt{V_{\rm rms}^2-V^2}$. In our JAM modeling, we integrated the MGE model (Table \ref{tab:MGEfit}) with a constant \ml$_{\rm F850LP} \approx 0.6$ \Msun/\Lsun\ (Section \ref{Sec:Stellar-mass-model}), isotropic stellar orbits ($\beta_z=0$), and an inclination angle of $i=65^\circ$. To assess the impact of a central IMBH on the simulated IFS and stellar kinematics, we considered two scenarios: (i) a model without a BH, and (ii) a model including an IMBH with a mass equivalent to 5\% of the NSC mass in FCC 119, corresponding to $M_{\rm BH} = 3.25\times10^5$ \Msun.

We generated the input-noiseless cubes for {\tt HSIM} within a FoV of $0\farcs4\times0\farcs4$ and a pixel size of $2\times2$ mas$^2$, following the below steps:

(i) We rebinned the MARCS SPS spectra with an assumed stellar population (Section \ref{sec:marcs}) onto a logarithmic scale using {\tt velscale} = 0.5 \kms, ensuring a constant $\Delta\log\lambda$ interval.

(ii) We constructed a Gaussian LOSVD kernel for each spatial position ($x,y$) with velocity $V(x,y)$ and velocity dispersion $\sigma_\star(x,y)$ using JAM, given specific dynamical parameters. The velocity was sampled at $\Delta V = 0.5$ \kms. 

(iii) We convolved the rebinned spectrum from step (i) with the Gaussian LOSVD from step (ii) in log-wavelength space, yielding the broadened stellar spectrum at position $(x,y)$.

(iv) The spectrum was rebinned onto a linear wavelength grid with a sampling step at least twice as fine as the HARMONI spectral resolution (e.g., K-long has $\Delta \lambda \approx 0.5$\AA). This process involved thorough integration across pixels to prevent any loss of information during \hsim\
simulating process.

(v) Finally, the spectrum at $(x,y)$ was then scaled to match the SB predicted by the MGE photometric model (Table \ref{tab:MGEfit}) for that position. This scaling factor was obtained by comparing the integrated flux of the original spectrum within the corresponding photometric band to the MGE model’s SB, using the {\tt ppxf.ppxf\_util.mag\_spectrum} function from the Penalized Pixel-Fitting ({\tt pPXF}) method\footnote{v9.2.1: \url{https://pypi.org/project/ppxf/}} \citep{Cappellari23}.

\subsubsection{{\hsim} output datacubes}\label{HSIM_outputs}

We used the input-noiseless cubes created in Section \ref{input-noiseless_cubes} as inputs for {\tt HSIM} to simulate HARMONI IFS observations in the $H$-high, $K$-short, and $K$-long gratings. The simulations were conducted under median observational conditions at the Armazones site, utilizing LTAO mode with a NGS of 17.5 mag in the $H$-band, located within a 30$\arcsec$ radial distance. We assumed an optical zenith seeing (0.5 $\mu$m) with a FWHM of $0\farcs64$ and an airmass of 1.3.

The exposure time for each simulation was carefully calibrated to achieve a minimum S/N of 3 per spaxel across the simulated FoV at the observed stellar features (before {\tt voronoi} binning). To replicate real-time observations, we incorporated multiple exposure frames and dithering, with a Detector Integration Time ({\tt DIT}) of 900 seconds per exposure. The total exposure time was determined by the number of exposures ({\tt NDIT}), following the relation {\tt DIT}$\times${\tt NDIT}.

\begin{figure}
\centering
\includegraphics[width=\textwidth]{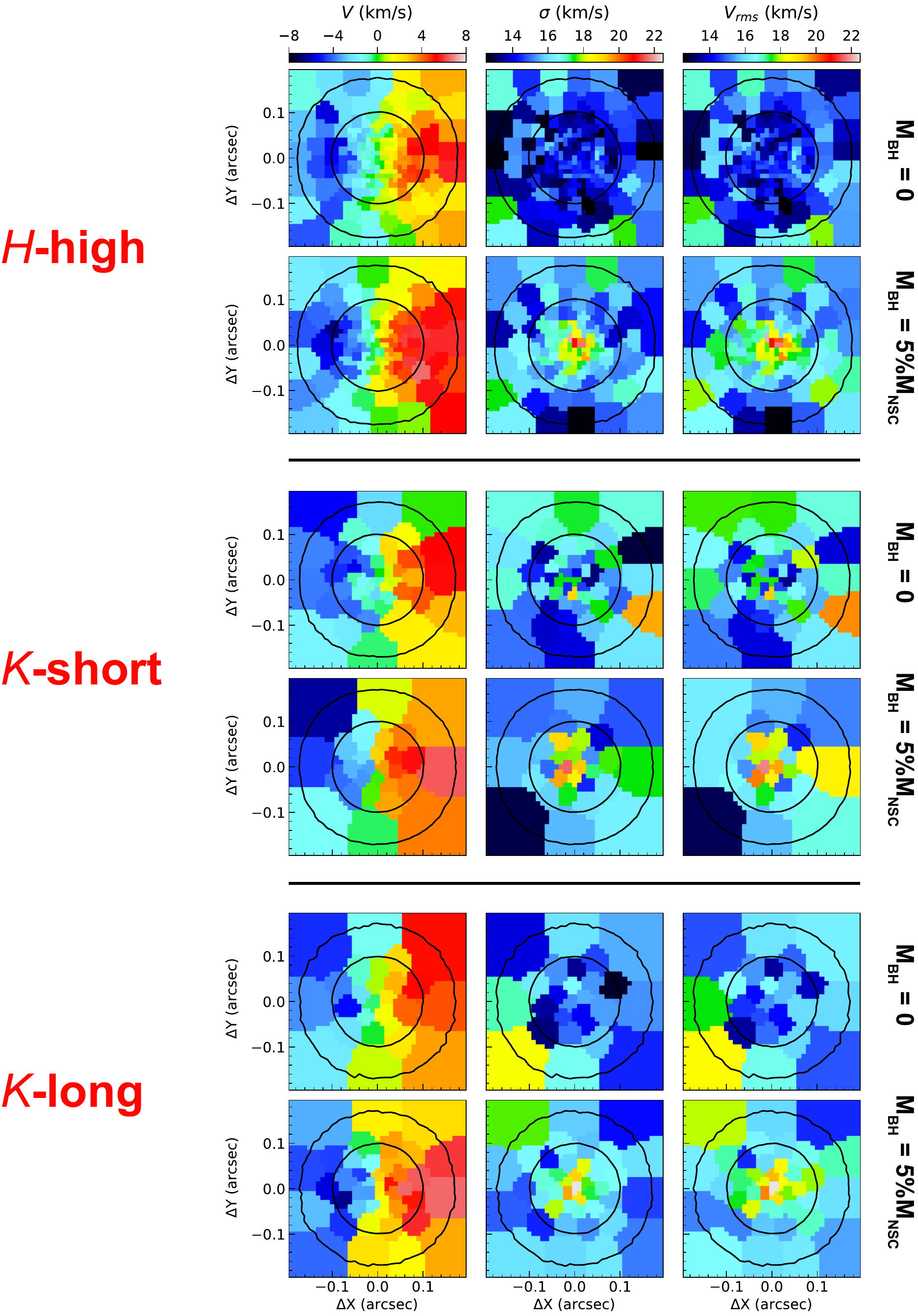}
\caption{The 2D stellar kinematic maps of FCC 119, derived from the three mock IFS gratings, show the velocity ($V$, left), velocity dispersion ($\sigma_\star$, middle), and root-mean-square velocity ($V_{\rm rms}$, right). The black contours in each map represent isophotes extracted from the collapsed mock IFS cubes, decreasing in brightness by 1 mag arcsec$^{-2}$ from the center outward. For each mock IFS grating, the kinematic maps were simulated assuming no BH (top) and including an IMBH (bottom).}
\label{fig:ppxf-extract-hhigh}
\end{figure}

\begin{figure}
    \begin{tabular}{cc}
        \textbf{{\textcolor{red}{M$_{\rm BH}=0$ ($H$-high)}}} & \textbf{{\textcolor{red}{M$_{\rm BH}$=5\% M$_{\rm NSC}$ ($H$-high)}}}\\
        \includegraphics[width=0.47\columnwidth]{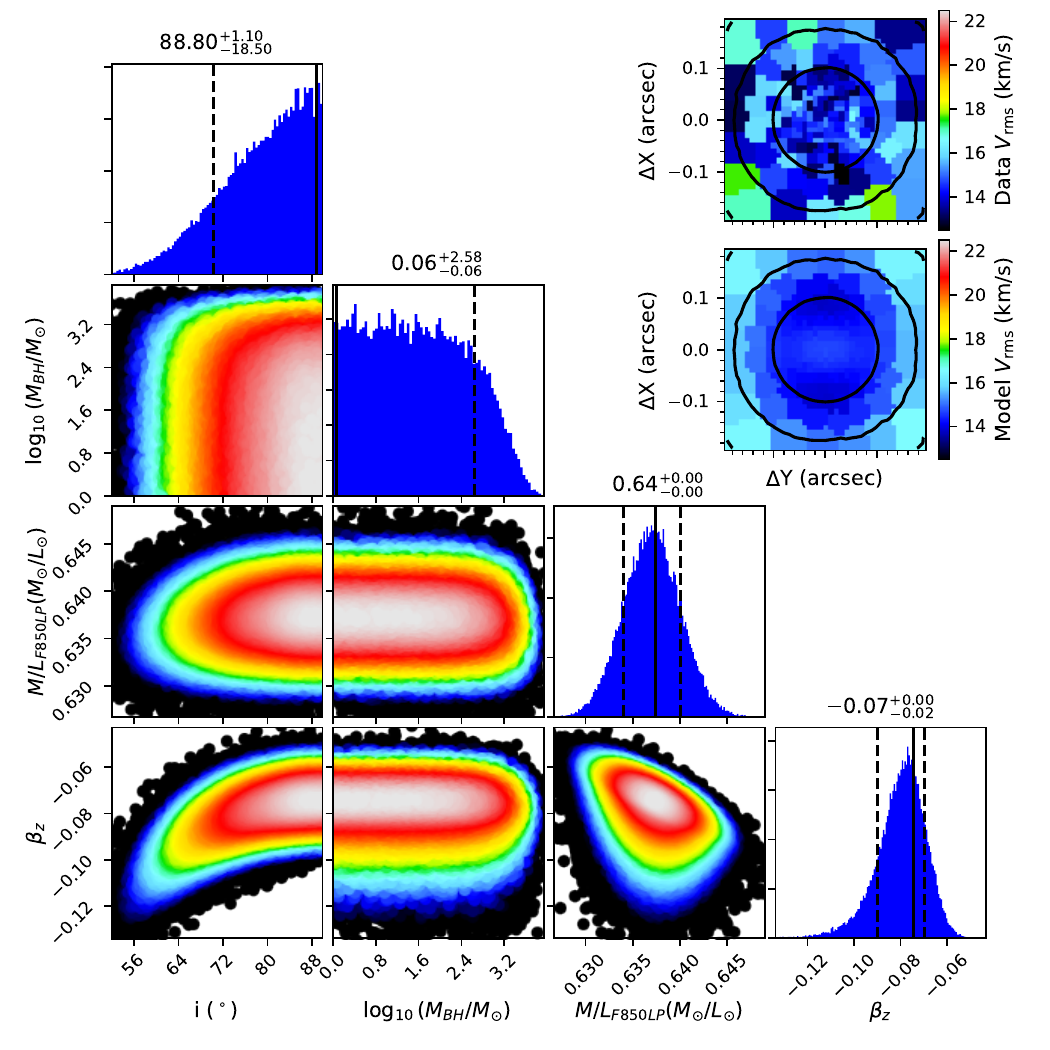} & \includegraphics[width=0.47\columnwidth]{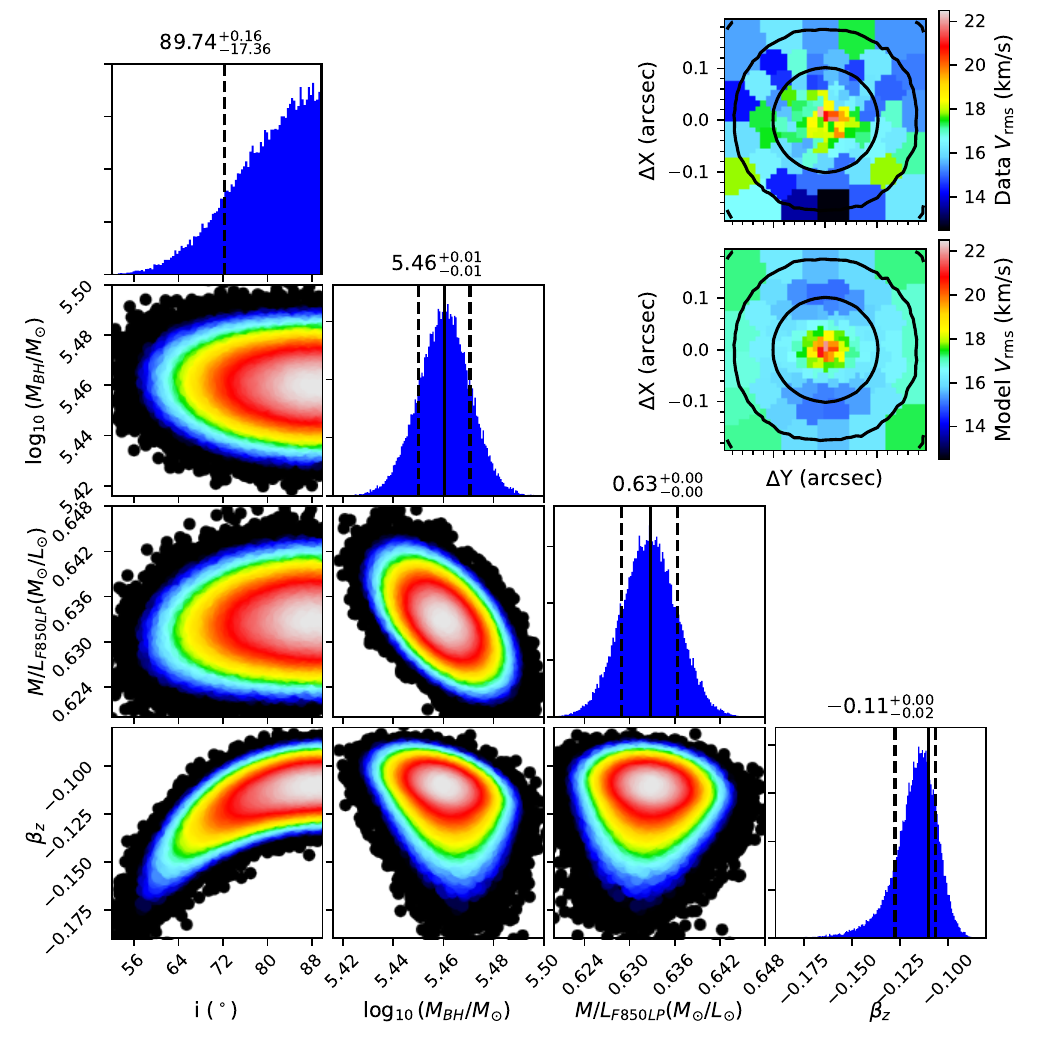} 
    \end{tabular}   
    \caption{The PDF of four parameters ($i$, $M_{\rm BH}$, $M/L_{\rm F850LP}$, $\beta_z$) constrained using JAM within the {\tt AdaMet} MCMC optimization framework, derived from mock $H$-high IFS data cubes. The figure presents results for two different \Mbh\ cases: no BH (left) and 5\%\Mnsc\ (right). Each data point in the 2D scatter plots represents a single model, with colors indicating the log-likelihood. The 1D histograms above display the corresponding projected 1D distributions. In the upper-right corner, a comparison is shown between the mock kinematic maps and their best-fitting model.} 
    \label{fig:adamet-hhigh}
\end{figure}

We obtained unexpected results regarding the exposure time required for the $K$-short grating to achieve stellar kinematics of comparable quality to the $K$-long. Given its higher flux level, the exposure time for $K$-short should theoretically be shorter than that for $K$-long. However, we found that $K$-short requires a longer exposure time (7.5 hours) compared to $K$-long (5 hours), arising because the $K$-short grating lacks strong stellar absorption features, such as the CO bandheads, which are present in the $K$-long range and are critical for detecting IMBH-induced stellar kinematics \citep{Nguyen2025b}. Furthermore, the atomic stellar features present in $K$-short are relatively weak and can be easily blended into spectral noise if the BH mass is large ($\gtrsim$$10^6$ \Msun) or if the exposure time is insufficient. This result highlights the limitations of the $K$-short IFS for detecting and accurately measuring the stellar kinematic signatures of central BHs.

\section{Results}\label{sec:discussion}

\subsection{Stellar kinematics extraction}\label{sec:kinematics}

We extracted the stellar LOSVD from the mock IFS data cubes generated in Section \ref{HSIM_outputs} using the adaptive {\tt Voronoi} binning technique via the {\tt vorbin} procedure\footnote{v3.1.5: \url{https://pypi.org/project/vorbin/}} \citep{Cappellari03} and the {\tt pPXF} method \citep{Cappellari23}. The {\tt Voronoi} binning optimally enhances the spectral S/N to a target threshold by setting {\tt targetSN = 25}, while {\tt pPXF} fits the binned spectra using the MARCS SPS templates (see Section \ref{sec:marcs}). The {\tt pPXF} fit was constrained to recover the velocity ($V$) and velocity dispersion ($\sigma_\star$) by setting {\tt moments = 2}. Since we did not account for the continuum and sky background, we excluded the additive and multiplicative polynomials by setting {\tt mdegree=0} and {\tt degree=-1}. The root-mean-square velocity was then computed as $V_{\rm rms}=\sqrt{V^2+\sigma_\star^2}$. During the fitting process, we accounted for the instrumental broadening of the HARMONI IFS by convolving the stellar templates with the constant instrumental dispersion difference between the data and the templates.

We enhanced the fidelity of our fits by incorporating 13 MARCS SPS templates with ages ranging from 3 to 15 Gyr and {\tt z004} metallicity. \citep{Nguyen2025b} demonstrated that using either the full grating spectrum or an optimally selected spectral range resulted in only a 5\% difference in the recovered kinematics. In this study, we selected the spectral ranges containing the strongest absorption features (as described in Section \ref{sec:createIFS} and shown in Figure \ref{fig:ppxf-spectrum}) to extract the stellar kinematics. The selected ranges are $\lambda$1.56--1.65 \micron, $\lambda$2.02--2.19 \micron, and $\lambda$2.21--2.36 \micron\ for the $H$-high, $K$-short, and $K$-long gratings, respectively.

The best-fit {\tt pPXF} models for the central bin, corresponding to the NSC center $(x,y) = (0,0)$, and the optimal spectral ranges of the $H$-high, $K$-short, and $K$-long data cubes are shown in Figure \ref{fig:ppxf-spectrum}. Figure \ref{fig:ppxf-extract-hhigh} further presents the resulting stellar LOSVD maps, which reveal significant differences in $\sigma_\star$ (and hence $V_{\rm rms}$) between the two scenarios: one without a BH and the other with an IMBH of \Mbh\ $=3.25\times10^5$ \Msun. In the absence of a BH, both $\sigma_\star$ and $V_{\rm rms}$ exhibit a central drop, whereas the IMBH models show a pronounced central peak in these kinematic quantities. This contrast represents a clear kinematic signature of a central massive object.

A central decrease in $\sigma_\star$ within resolved nuclei is a common feature in systems that either lack a central BH or host a low-mass BH \citep{Nguyen2025b}. This behavior occurs across a broad range of Sérsic indices ($n$) and power-law indices ($\gamma$) in the core-Sérsic profile of an NSC, as theoretically predicted by \citep{Tremaine94} and numerically demonstrated by \citep{Nguyen2025b} using HARMONI IFS stellar kinematics simulations. In contrast, the presence of an IMBH induces a central increase in $\sigma_\star$ due to the IMBH’s gravitational dominance within its SOI. The difference in $V_{\rm rms}$ between scenarios with and without an IMBH is $\approx$6 \kms, corresponding to 30\% of the nuclear $\sigma_\star$ of FCC 119 ($\sigma_\star \approx 20$ \kms) \citep{Ding2023, Iodice2019a}. This difference is substantial enough to reliably distinguish between the two cases.

\begin{figure}
    \begin{tabular}{cc}
        \textbf{\Large{\textcolor{red}{M$_{\rm BH}=0$ ($K$-short)}}} & \textbf{\Large{\textcolor{red}{M$_{\rm BH}$=5\% M$_{\rm NSC}$ ($K$-short)}}}\\
        \includegraphics[width=0.47\columnwidth]{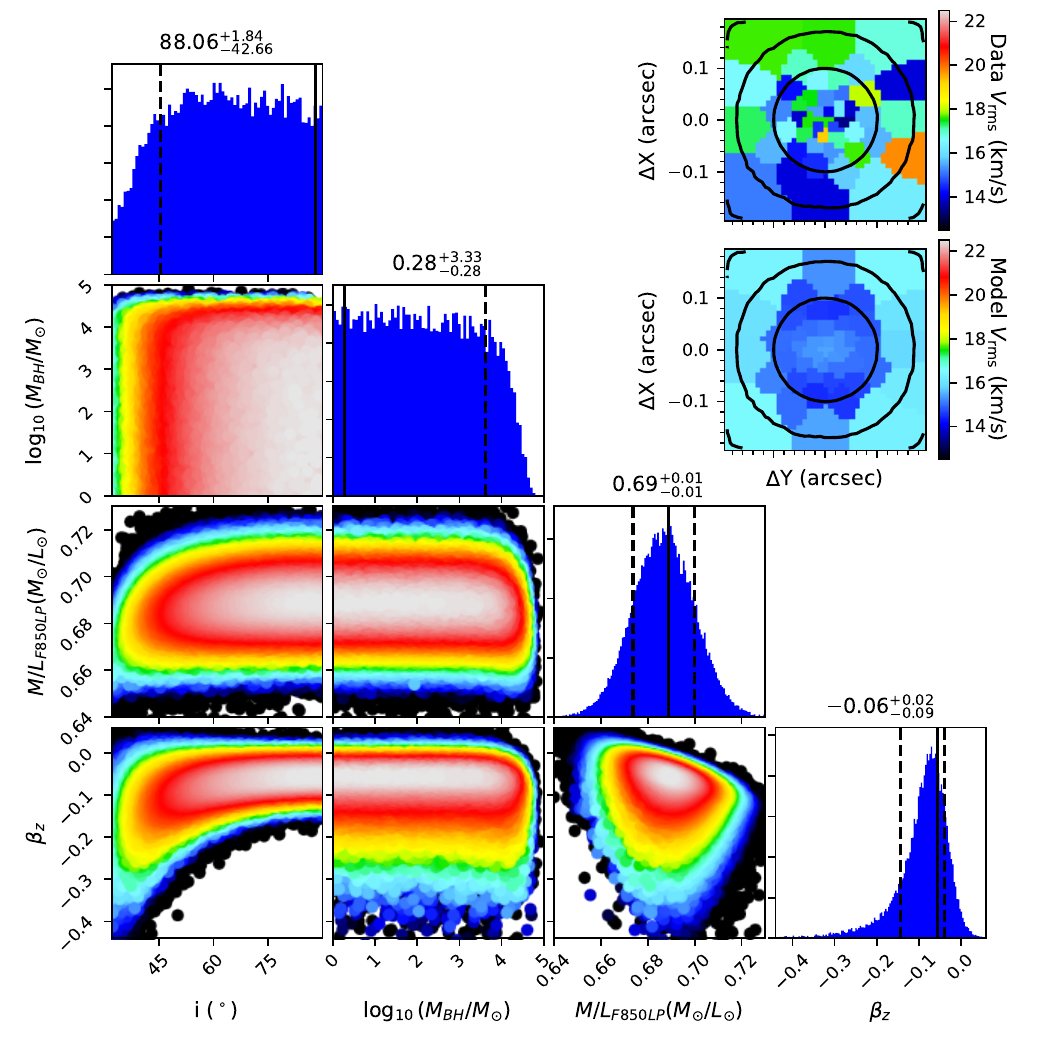} & \includegraphics[width=0.47\columnwidth]{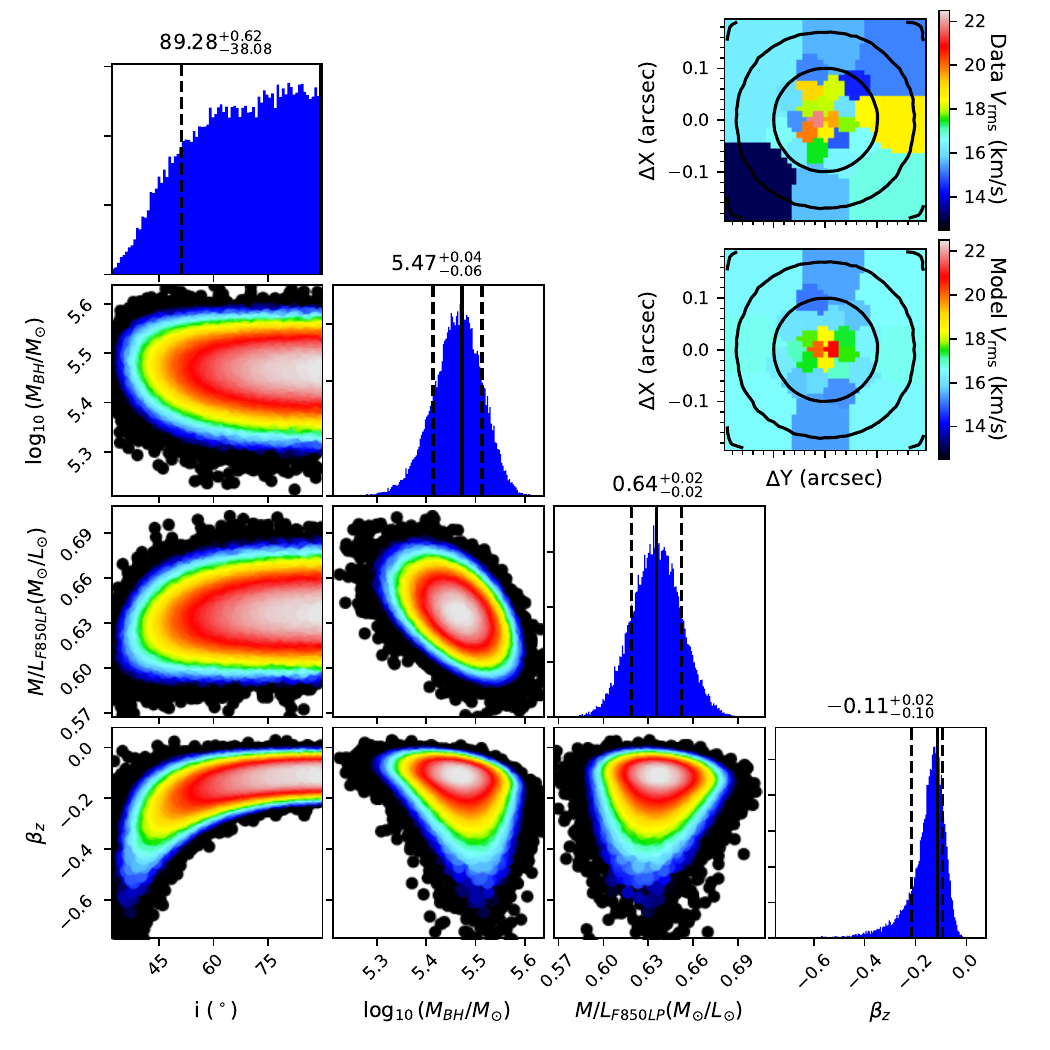} \vspace{2mm} \\
        
        \textbf{\Large{\textcolor{red}{M$_{\rm BH}=0$ ($K$-long)}}} & \textbf{\Large{\textcolor{red}{M$_{\rm BH}$=5\% M$_{\rm NSC}$ ($K$-long)}}}\\
        \includegraphics[width=0.47\columnwidth]{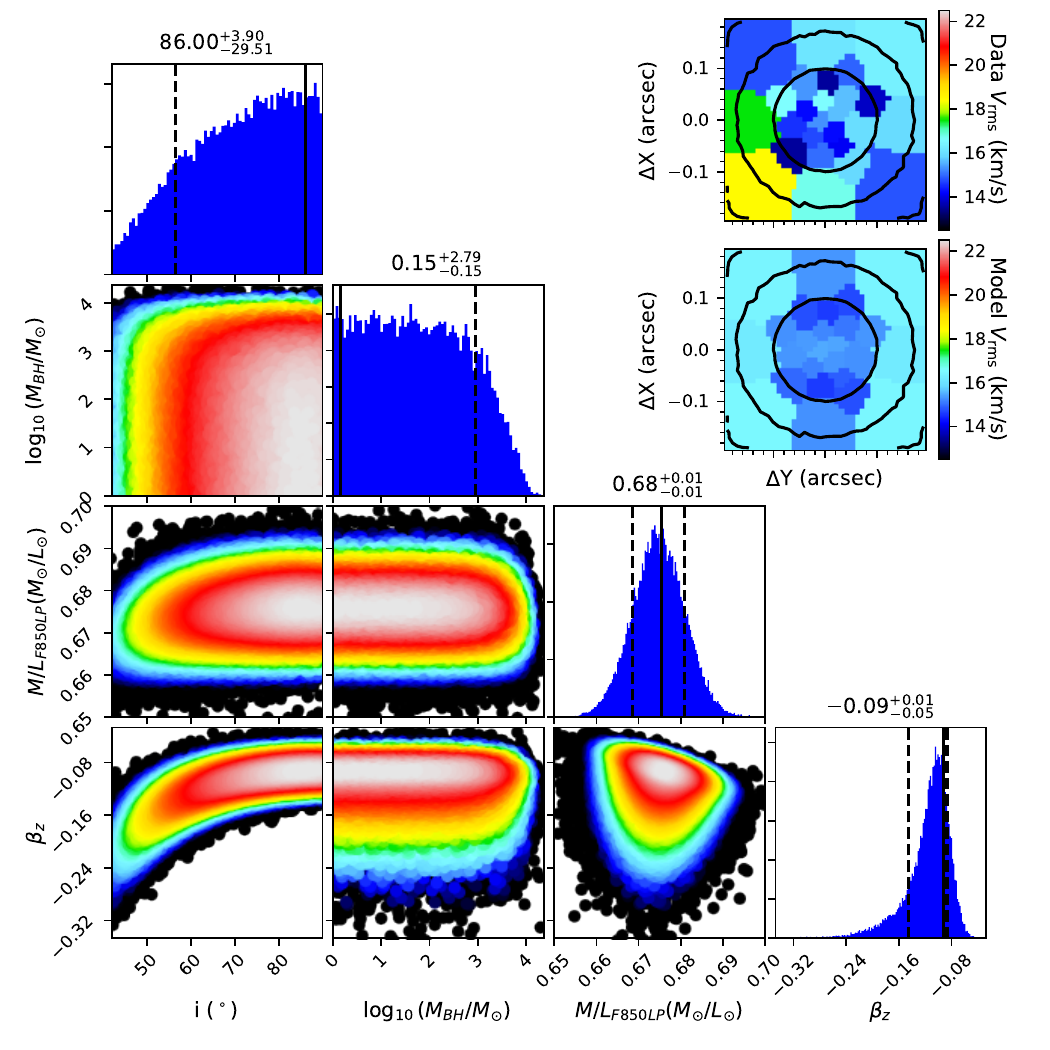} & \includegraphics[width=0.47\columnwidth]{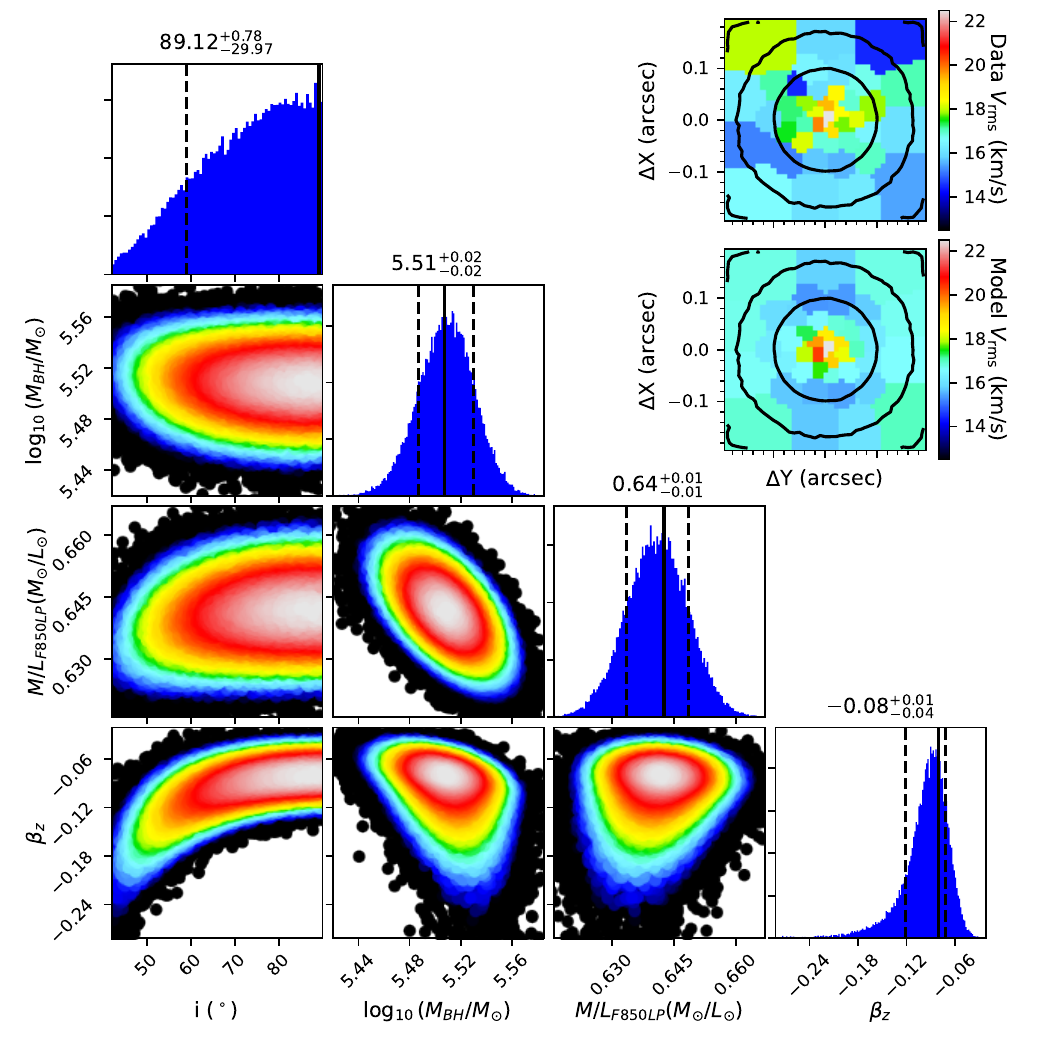} 
    \end{tabular}
    \caption{Same as Figure \ref{fig:adamet-hhigh}, but for the $K$-short and $K$-long gratings.}
    \label{fig:adamet-kshort}
\end{figure}

\begin{table*}
\centering\footnotesize
\caption{Best-fit JAM models and uncertainties.}
\begin{adjustwidth}{-\extralength}{0cm}
\centering
\begin{tabular}{cc|ccc|ccc}
\hline\hline
 Initial & Input & \multicolumn{3}{c|}{$M_{\rm BH}=0$} & \multicolumn{3}{c}{$M_{\rm BH}=5\%\,M_{\rm NSC}$} \\
\cline{3-8}
parameters & JAM guess& best-fit & (16--84)\% & (0.14--99.86)\% & best-fit & (16--84)\% & (0.14--99.86)\% \\
\hline
(1) & (2) & (3) & (4) & (5) & (6) & (7) & (8)  \\
\hline
&&\multicolumn{6}{c}{$H$-high} \\
\hline
$i$ ($^\circ$) & 65 & 89 & $>$19  & $>$36 & 90 & $>$17  & $>$34  \\ 
$\log_{10}(M_{\rm BH}$/M$_\odot)$ & (*) & 0.06 & $<$2.58 & $<$3.68 & 5.460 & $\pm$0.010 & $\pm$0.031 \\ 
$M/L_{\rm F850LP}$ (M$_\odot$/L$_\odot$) & 0.6 & 0.637 & $\pm$0.003 & $\pm$0.009 & 0.633 & $\pm$0.004 & $\pm$0.011 \\ 
$\beta_z$ & 0 & $-$0.074 & $\pm$0.010 & $\pm$0.039 & $-$0.110 & $\pm$0.011 & $\pm$0.044 \\ 
\hline
&&\multicolumn{6}{c}{$K$-short} \\
\hline
$i$ ($^\circ$) & 65 & 88 & $>$43 & $>$56 & 89 & $>$38 & $>$56 \\ 
$\log_{10}(M_{\rm BH}$/M$_\odot)$ & (*) & 0.28 & $<$3.33 & $<$4.37 & 5.472 & $\pm$0.050 & $\pm$0.150 \\ 
$M/L_{\rm F850LP}$ (M$_\odot$/L$_\odot$) & 0.6 & 0.689 & $\pm$0.013 & $\pm$0.039 & 0.636 & $\pm$0.017 & $\pm$0.050 \\ 
$\beta_z$ & 0 & $-$0.057 & $\pm$0.052 & $\pm$0.240 & $-$0.114 & $\pm$0.062 & $\pm$0.340 \\ 
\hline
&&\multicolumn{6}{c}{$K$-long} \\
\hline
$i$ ($^\circ$) & 65 & 86 & $>$30 & $>$48 & 89 & $>$30 & $>$50 \\ 
$\log_{10}(M_{\rm BH}$/M$_\odot)$ & (*) & 0.15 & $<$2.79 & $<$3.92 & 5.507 & $\pm$0.022 & $\pm$0.065 \\ 
$M/L_{\rm F850LP}$ (M$_\odot$/L$_\odot$) & 0.6 & 0.675 & $\pm$0.006 & $\pm$0.019 & 0.643 & $\pm$0.008 & $\pm$0.022 \\ 
$\beta_z$ & 0 & $-$0.093 & $\pm$0.029 & $\pm$0.130 & $-$0.080 & $\pm$0.025 & $\pm$0.230 \\ 
\hline
\end{tabular}
\vspace{2mm}
\parbox[t]{\textwidth}{\textit{Notes:} Columns 1--2: The free parameters of JAM, their initial guessed values. The asterisk (*) in \Mbh\ in Column 2 indicates that the initial guesses for the \Mbh\ either zero or \Mbh\ $=3.25 \times 10^5$~\Msun.  Columns 3--5: the best-fitting parameters for the case without a central BH (\Mbh $=0$ \Msun), their 1$\sigma$ and 3$\sigma$ uncertainties. Columns 6--8: Same as Columns 4--5 but for the input \Mbh\ $=3.25 \times 10^5$~\Msun.}
\label{tab:adamet-value}
\end{adjustwidth}
\end{table*}

\subsection{The black hole mass recovering}\label{sec:mbhcovering}

We combined the MGE model (Section \ref{Sec:mge}) and the mock stellar kinematic measurements (Section \ref{sec:kinematics}) to reconstruct the LOSVD of FCC 119, including $V$ and $\sigma_\star$ (or $V_{\rm rms}$), using JAM (Section \ref{sec:jam}). The fitting was performed within a Bayesian inference framework \citep{Cappellari13a}, employing a Markov Chain Monte Carlo (MCMC) simulation via the adaptive Metropolis algorithm implemented in {\tt AdaMet}\footnote{v2.0.9: \url{https://pypi.org/project/adamet/}} \citep{Haario01}. We assumed a uniform Gaussian error distribution, making the posterior probability proportional to the logarithm of the likelihood, $\ln{P(\rm data|model)} \propto -0.5\chi^2$, where the chi-squared $\chi^2$ is given by:
\begin{equation} 
\chi^2 \equiv \sum_i \left(\frac{V_{{\rm rms},i} - \overline{V}_{{\rm rms},i}}{\sigma_i}\right)^2,
\end{equation}
Here, $V_{{\rm rms},i}$ and $\sigma_i$ represent the root-mean-square velocity and its associated error for bin $i$ extracted from the mock data. The model root-mean-square velocity, $\overline{V}_{{\rm rms},i}$, was computed using JAM and convolved with the LTAO PSF, which has a FWHM$_{\rm PSF}=0\farcs012$ \citep{Thatte20, Nguyen23, Nguyen2025b}. The JAM model explores a four-parameter space, including the black hole mass ($M_{\rm BH}$) in logarithmic scale, as well as the inclination ($i$), the mass-to-light ratio in the F850LP filter ($M/L_{\rm F850LP}$), and the anisotropy parameter ($\beta_z$) in linear scale.

We performed a total of $3 \times 10^5$ calculations and excluded the first 20\% of the MCMC steps as the burn-in phase. The best-fit parameters and their uncertainties were then derived from the probability distribution function (PDF) of the remaining 80\% of the calculations. We used the same initial guesses for the free parameters in all fittings, which are: $i = 65\degr$, $M/L_{\rm F850LP} = 0.6$ (M$_\odot$/L$_\odot$), $\beta_z = 0$, and $M_{\rm BH}$ either 0~\Msun\ or $3.25 \times 10^5$\Msun, identical to the input values for {\tt HSIM}. We also defined the search ranges as follows: $i$ (33\degr–90\degr), $\log_{10} M_{\rm BH}$ (0–7\Msun), $M/L_{\rm F850LP}$ (0–2~M$\odot$/L$\odot$), and $\beta_z$ ($-1$ to 0.99)

We present the {\tt AdaMet} MCMC results for the $H$-high, $K$-short, and $K$-long bands, corresponding to both input $M_{\rm BH}$ values, in Figures \ref{fig:adamet-hhigh} and \ref{fig:adamet-kshort}. The 2D distribution of each parameter pair is shown in the corner plots after marginalizing over the other parameters, where each point represents a JAM model. The color of these scatter points, ranging from white to blue, indicates their likelihood at different confidence levels (CLs), from $1\sigma$ to $3\sigma$. Black points represent CLs greater than $3\sigma$, while white points denote the highest likelihood (best fit) within the $1\sigma$ CL. The top histogram illustrates the 1D distribution of each parameter and is used to estimate the best-fit values and associated uncertainties. An anti-correlation between $M/L_{\rm F850LP}$ and $M_{\rm BH}$ is evident in their 2D PDFs, arising from the interplay between the gravitational potentials of the IMBHs and their host galaxies, where larger BHs correspond to lower $M/L_{\rm F850LP}$ values and vice versa. This behavior highlights the high spatial resolution of our simulations, enabled by the 10 mas observational scale, which is sufficient to resolve the SOIs of these IMBHs at the distance of the Fornax Cluster.

The top-right corner of these figures compares the $V_{\rm rms}$ map extracted from the \hsim\ datacubes (top) with that derived from the best-fit JAM model (bottom), both using the same color scale. The best-fit parameter values, along with their associated $1\sigma$ and $3\sigma$ uncertainties, are listed in Table \ref{tab:adamet-value}. Our recovered values for both $M/L_{\rm F850LP}$ and $M_{\rm BH}$ closely match the input values, with differences of $<$15\%. Notably, the uncertainties from the MCMC fits remain relatively small ($<$10\% at the $3\sigma$ CL). This precision is attributed to the exceptional spatial and spectral resolutions of our IFS simulations, which effectively resolve the BH’s SOI and capture subtle increasing in $V_{\rm rms}$ induced by the presence of a small IMBH.

In addition, for a scenario without a central IMBH ($M_{\rm BH}=0$), the MCMC yields an upper limit of $M_{\rm BH}\approx10^4$ \Msun. This indicates HARMONI’s ability to detect IMBHs above this mass threshold at a distance of 20 Mpc. Conversely, simulations with an IMBH of $M_{\rm BH} \approx 3\times10^5$ \Msun\ and $\sigma_\star = 20$ \kms\ at a distance of 20 Mpc give an \Rsoi~$\approx0\farcs032$. This radius is three times larger than the simulated spaxel size of $0\farcs01\times0\farcs01$, demonstrating that HARMONI can resolve IMBHs with masses as low as $M_{\rm BH} \approx 10^5$ \Msun\ at 20 Mpc. This establishes a fundamental lower limit for $M_{\rm BH}$ measurements using dynamical techniques with ELT/HARMONI in the future.

The recovered $\beta_z$ values exhibit a broad range from $-0.5$ to $-0.03$, significantly deviating from the input value of $\beta_z = 0$. This suggests that stars within the nucleus of FCC 119 predominantly follow tangential orbits. These discrepancies may arise from the scattered nature of the $V_{\rm rms}$ measurements at the FoV edges. The faintness of FCC 119, combined with a steep decline in S/N toward the FoV boundaries, affects the accuracy of spectral fitting in these regions.

Additionally, the inclination $i$ is not well constrained, varying between $33^{\circ}$ and $90^{\circ}$. This is likely due to the kinematic properties of FCC 119, where velocity dispersion dominates and rotational motion is minimal ($V \approx \pm 8$ \kms). As discussed in \citep{Nguyen2025b}, NSCs with little to no rotation exhibit kinematic characteristics similar to nearly spherical systems. In such cases, galaxies appear similar over a wide range of inclinations, making it difficult to precisely constrain $i$ in our models.

\begin{figure}
    \centering
    \includegraphics[width=0.7\columnwidth]{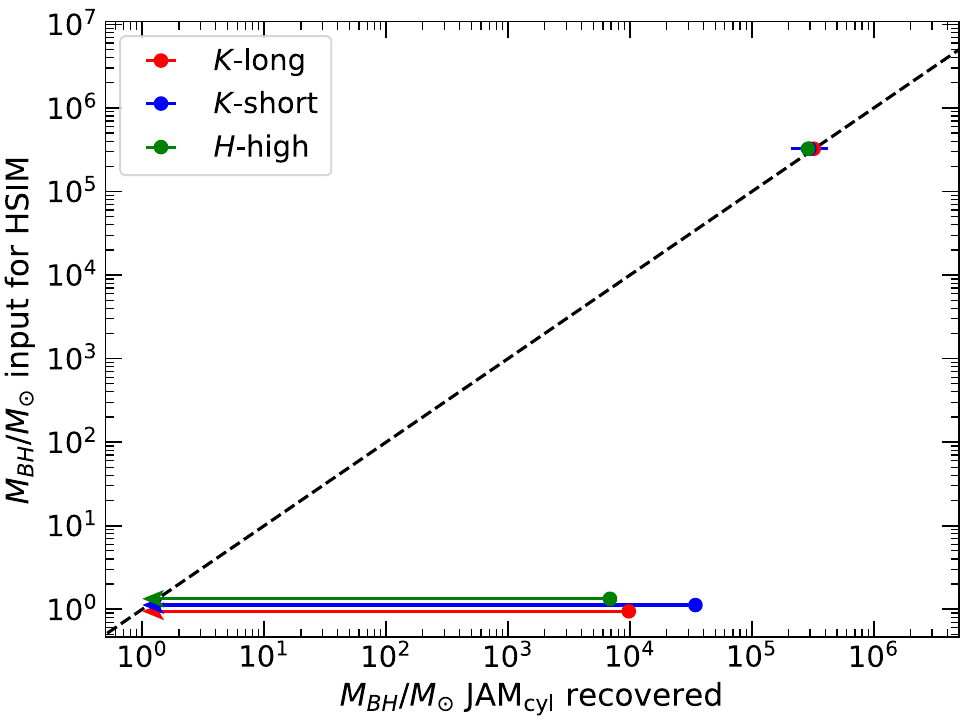}
    \caption{A comparison of the input \Mbh\ used in \textsc{HSIM} and the recovered values within $3\sigma$ CL. The black dashed line represents the line of equality between the input \Mbh\ in \textsc{HSIM} and our recoveries.}
    \label{fig:BH-recovery}
\end{figure}

We compared the input $M_{\rm BH}$ values used to generate the mock IFS data cubes with the corresponding recovered values in Figure \ref{fig:BH-recovery}. The recovered data points closely align with the line of equality, demonstrating the robustness of our method for constraining IMBH masses using HARMONI observations.

\subsection{Dynamical mass-segregation in FCC 119 NSC?}\label{mass-segregation}

In a stellar system, stars will reach an energy equipartition state through two-body interactions, causing higher-mass stars decrease velocities and move toward the center of the system, while lower-mass stars gain higher velocities and drift outward (i.e., mass-segregation process) \citep{Spitzer1987}. Such massive star's deaths leads to a gradual accumulation of dark remnants (e.g., neutron stars, stellar-mass black holes) and cause an increase in the \ml\ toward the central region \citep{Bianchini17}. The central peak-up of \ml~ in a star cluster system can produce a stellar kinematics that mimic the presence of a central IMBH and potentially leading to an ambiguous IMBH detection \citep{Nguyen2025b}. 

\begin{figure}
    \centering
    \includegraphics[width=0.7\columnwidth]{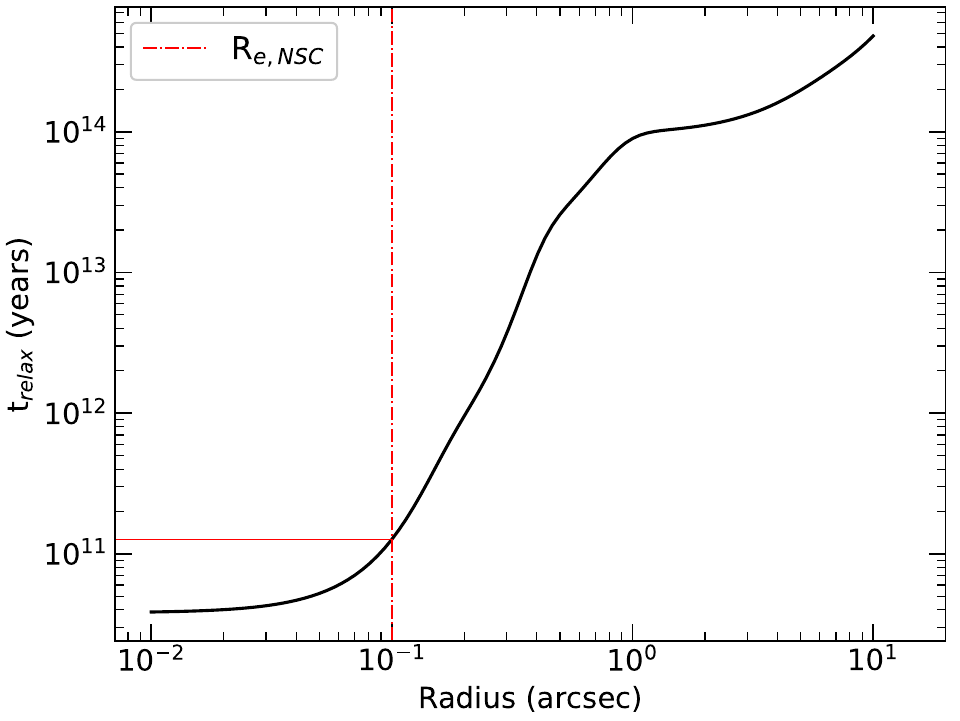}
    \caption{Relaxation time (\trelax) as a function of radius plots along the semi-major axis of FCC 119 (black-solid line). The red dash-dotted line indicates the effective radius of its NSC ($R_{\rm e, NSC}$)}
    \label{fig:t-relax}
\end{figure}

To assess the impact of this mass-segregation phenomenon on the FCC 119's NSC, we evaluated the relaxation time (\trelax) as a function of radius \citep{Bahcall77, Valluri05}:
\begin{equation}
\label{eq:t-relax}
\centering
	t_{\rm relax}({\emph r}) \approx (1.4\times10^8 \mathrm{yr})\,\sigma^3_{20}(r)\,\rho_{5}^{-1}(r)(\ln\Lambda_{10})^{-1},
\end{equation}
where $\sigma_{20}\equiv\sigma_\star/20$ \kms,  $\rho_{5}\equiv\rho/10^5$ \Msun\  pc$^{-3}$, $\ln\Lambda_{10} \equiv\ln(\Lambda)/10$ ($\ln\Lambda=10$). We also used a constant average $\sigma_\star=20$ \kms~\citep{Ding2023, Iodice2019a}. The radial mass density $\rho(r)$ of FCC 119 is estimated using the HST/WFC F850LP MGE model (Table \ref{tab:MGEfit}) with a $M/L_{\rm F850LP}\approx0.6\ ({\rm M_\odot/L_\odot}$). We plotted in Figure \ref{fig:t-relax} the relaxation time \trelax~as a radial function, with the value at the effective radius $R_{\rm e,NSC}=0\farcs1$ (Section \ref{Sec:Stellar-mass-model}) of $t_{\rm relax}(R_{\rm e, NSC})\approx1.5\times10^{11}$ year. This timescale is significantly longer than the age of the universe. It suggests that the NSC of FCC 119 might not undergo a significant migration of dark remnants. Consequently, we can disregard the mass-segregation process in FCC 119 and confidently attribute the central kinematic peak to the gravitational potential of an IMBH but see \citep{Nguyen2025b} for a specific case.

\subsection{Constraint on nearby NSC brightness and sensitivity}\label{limit}

We examined the sensitivity of the ELT to low-SB galaxies, focusing on FCC 119, located at a distance of 20 Mpc, which has a SB of 20.1 mag arcsec$^{-2}$ at a radius of 1\arcsec\ in the $z$-band \citep{Turner2012}. To assess this, we generated mock high-spectral-resolution HARMONI IFS cubes with varying exposure times to determine the minimum integration time required to achieve a sufficient S/N ratio for robust kinematic LOSVD measurements with each grating. Our simulations indicate that the required exposure times are at least 3, 6, and 4 hours for the $H$-high, $K$-short, and $K$-long bands, respectively, as summarized in Table \ref{table:hsim-exptime}. These sensitivity estimates account only for the on-source science time and do not include additional time for target acquisition, instrument overheads, and LTAO setup. Consequently, the total required integration time in practice is expected to be at least twice as long. Furthermore, the SB profile was derived from HST imaging and extrapolated down to a spatial scale of $0\farcs01$ toward the galaxy’s central region, potentially necessitating longer exposure times than initially estimated.

Thus, we established a limiting distance of 20 Mpc for conducting HARMONI observations within a reasonably short exposure time. At this distance, the NSC must have a central SB as low as $\mu_{\rm c} \approx 16$ mag arcsec$^{-2}$, decreasing to approximately 18 mag arcsec$^{-2}$ at a radius of $0\farcs2$ (see Figure \ref{fig:IRAFextrapolate}). \citep{Nguyen2025b} reported a similar SB decline in the NSCs of two much closer galaxies, NGC 300 and NGC 3115 dw01, where the central SB is $\mu_{\rm c} \approx 15$ mag arcsec$^{-2}$, dropping to about 17 mag arcsec$^{-2}$ at a radius of $0\farcs2$. This consistency reinforces our observational lower limit. Notably, our findings indicate a sensitivity approximately two orders of magnitude lower than previous limits for 8–10 meter class telescopes, which are capable of observing nuclei with a central SB above $\mu_{\rm c} \approx 14$ mag arcsec$^{-2}$, decreasing to about 16 mag arcsec$^{-2}$ at a radius of $1\farcs5$.

\section{Conclusions}\label{sec:Conclusions}

We expanded the current HARMONI IMBH sample, which was limited within 10 Mpc \citep{Nguyen2025b} to 20 Mpc by exploring the capabilities of HARMONI for dynamically measuring the masses of IMBHs in the second faintest dwarf and nucleated member of the Fornax Cluster, FCC 119, which is also much fainter than almost all members of the Virgo Cluster. Based on our findings, our conclusions are summarized as follows:

(i) We compiled an expanded sample of 85 dwarf galaxies hosting bright NSCs, with masses ranging from $1.6 \times 10^5$ to $2.3 \times 10^8$ \Msun\ and effective radii between 0.5 and 62 pc, located at distances of 10–20 Mpc, representing promising sites for IMBHs.

(ii) The sample spans a variety of galaxy types, including ellipticals (27\%), lenticulars (29\%), spirals (39\%), and irregulars (5\%). It comprises 29 members of the Virgo Cluster, 20 members of the Fornax Cluster, and 36 isolated galaxies.

(iii) We performed HARMONI simulations for FCC 119 using the $H$-high, $K$-short, and $K$-long gratings with a spaxel scale of $0\farcs01\times0\farcs01$, requiring on-source exposure times of 4, 7.5, and 5 hours, respectively. The consistent stellar kinematic measurements obtained across the mock IFS data cubes validate our methodology and observational strategy, demonstrating the reliability of extending the IMBH sample and highlighting HARMONI’s capabilities.

(iv) The recovered $M_{\rm BH}$ and stellar $M/L_{\rm F850LP}$ closely match the input simulation values, with uncertainties $\lesssim$5\%, demonstrating the robustness and reliability of our measurement technique using HARMONI data.

(v) We determined that HARMONI can effectively observe NSCs at the distance of the Fornax Cluster with a central surface brightness as low as $\mu_c \approx 16$ mag arcsec$^{-2}$ (decreasing to $\sim$18 mag arcsec$^{-2}$ at $0\farcs2$) and can dynamically resolve IMBHs with masses down to $\sim$$10^5$ \Msun.

\section*{Acknowledgements}

This research is funded by University of Science, VNU-HCM under grant number T2023-105. This research is based on observations made with the NASA/ESA Hubble Space Telescope, and obtained from the Hubble Legacy Archive, which is a collaboration between the Space Telescope Science Institute (STScI/NASA), the Space Telescope European Coordinating Facility (ST-ECF/ESA) and the Canadian Astronomy Data Centre (CADC/NRC/CSA). This research made use of hips2fits,\footnote{https://alasky.cds.unistra.fr/hips-image-services/hips2fits} a service provided by CDS.

{\it Facilities:} \hst, DESI, SDSS, 2MASS

{\it Software:} {\tt Python~3.12:} \citep{VanRossum2009}, 
{\tt Matplotlib~3.6:} \citep{Hunter2007}, 
{\tt NumPy~1.22:} \citep{Harris2020}, 
{\tt SciPy~1.3:} \citep{Virtanen2020},  
{\tt photutils~0.7:} \citep{bradley2024}, 
{\tt AstroPy~5.1} \citep{AstropyCollaboration22}, 
{\tt AdaMet 2.0} \citep{Cappellari13a}, 
{\tt JamPy~7.2} \citep{Cappellari20}, 
{\tt pPXF~8.2} \citep{Cappellari23}, 
{\tt vorbin~3.1} \citep{Cappellari03}
{\tt MgeFit~5.0} \citep{Cappellari02}, and
{\tt HSIM 3.11} \citep{Zieleniewski15}.


\section*{Data Availability}

All data and software used in this paper are public. We provided their links in the text when discussed.

\reftitle{References}


\bibliography{references}

\PublishersNote{}
\end{document}